\documentclass[a4paper,twoside]{article}
\usepackage{amsmath,amssymb}
\usepackage{epsfig,rotating}

\newlength{\gfxlabelwidth}

\newcommand{\wlabelgfx}[4]{
    \settowidth{\gfxlabelwidth}{\begin{sideways}$#3$\end{sideways}}
    \begin{tabular}[c]{@{}c@{\hspace{2mm}}c@{}}
      \parbox{\gfxlabelwidth}{\begin{sideways}$#3$\end{sideways}}&
      \parbox{#4}{\epsfig{file=#1,width=#4}}            \\
      & $#2$
    \end{tabular} }

\newcommand{\pabl}[2]{\frac{\partial #1}{\partial #2}}
 
\newcommand{\ppabl}[3]{\frac{\partial^2#1}{\partial#2\partial#3}}
\newcommand{\PPabl}[2]{\frac{\partial^2#1}{\partial #2^2}}

\newcommand{\anf}[1]{``#1''}

\newcommand{\sign}{\operatorname{sign}}
\newcommand{\Tr}{\operatorname{Tr}}
\newcommand{\Jac}{\operatorname{Jac}}
\newcommand{\Hess}{\operatorname{Hess}}
\newcommand{\cbrt}[1]{\sqrt[\leftroot{3}\uproot{3} 3]{#1}}
\newcommand{\ord}{} 

\renewcommand{\vec}[1]{{\bf #1}}
\renewcommand{\Re}{{\rm Re}\,}
\renewcommand{\Im}{{\rm Im}\,}

\newcommand{\Z}{\mathbb{Z}}
\newcommand{\cH}{{\cal H}}
\newcommand{\cF}{{\cal F}}

\begin{document}
\title{Uniform approximations for non-generic bifurcation scenarios
  including bifurcations of ghost orbits}     
\author{T Bartsch, J Main, and G Wunner\\ 
        Institut f\"ur Theoretische Physik und Synergetik \\
        Universit\"at Stuttgart \\
        D-70550 Stuttgart, Germany}

\maketitle

\begin{abstract}
  Gutzwiller's trace formula allows interpreting the density of states
  of a classically chaotic quantum system in terms of classical
  periodic orbits. It diverges when periodic orbits undergo
  bifurcations, and must be replaced with a uniform approximation in
  the vicinity of the bifurcations. As a characteristic feature, these
  approximations require the inclusion of complex ``ghost orbits''. By
  studying an example taken from the Diamagnetic Kepler Problem,
  viz. the period-quadrupling of the balloon-orbit, we demonstrate
  that these ghost orbits themselves can undergo bifurcations, giving
  rise to non-generic complicated bifurcation scenarios. We extend
  classical normal form theory so as to yield analytic descriptions of
  both bifurcations of real orbits and ghost orbit bifurcations. We
  then show how the normal form serves to obtain a uniform
  approximation taking the ghost orbit bifurcation into account. We
  find that the ghost bifurcation produces signatures in the
  semiclassical spectrum in much the same way as a bifurcation of real
  orbits does.
\end{abstract}

\newpage

\section{Introduction}

In the ``old'' quantum theory developed around the turn of the
century, quantization of a mechanical system used to be based on its
classical behavior. In 1917, Einstein \cite{Ein17} was able to
formulate the quantization conditions found by Bohr and Sommerfeld in
their most general form. At the same time, however, Einstein pointed
out that they were applicable only to systems whose classical phase
space was foliated into invariant tori, i.e., which possess
sufficiently many constants of motion, and that most mechanical
systems do not meet this requirement. The development of quantum
mechanics by Schr\"odinger, Heisenberg and others then offered
techniques which allowed for a precise description of atomic systems
without recourse to classical mechanics. Thus, the problem of
quantizing chaotic mechanical systems on the basis of their classical
behavior remained open.

As late as in the 1960s, Gutzwiller returned to what is now known as a
semiclassical treatment of quantum systems. Starting from Feynman's
path integral formulation of quantum mechanics, he derived a
semiclassical approximation to the Green's function of a quantum
system which he then used to evaluate the density of states. His
\emph{trace formula} \cite{Gut71,Gut90} is the only general tool known
today for a semiclassical understanding of systems whose classical
counterparts exhibit chaotic behavior. It represents the quantum
density of states as a sum of a smooth average part and fluctuations
arising from all periodic orbits of the classical system, and
therefore allows structures in the quantum spectrum to be interpreted
in terms of classical mechanics. The derivation of the trace formula
assumes all periodic orbits of the system to be isolated. Thus, it is
most appropriate for a description of completely hyperbolic system,
where in some cases it even allows for a semiclassical determination
of individual energy levels, as was done, e.g., by Gutzwiller in the
case of the Anisotropic Kepler Problem \cite{Gut82}. In generic
Hamiltonian systems exhibiting mixed regular-chaotic dynamics,
however, bifurcations of periodic orbits can occur. They cause the
trace formula to diverge because close to a bifurcation the periodic
orbits involved approach each other arbitrarily closely.

This failure can be overcome if all periodic orbits involved in a
bifurcation are treated collectively. A first step in this direction was
taken by Ozorio de Almeida and Hannay \cite{Alm87} who proposed formulas
for the collective contributions which yield finite results at the
bifurcation energy but do not correctly reproduce the results of
Gutzwiller's trace formula as the distance from the bifurcation increases.
Similarly, Peters, Jaff\'e and Delos \cite{Pet94} were able to deal with
bifurcations of closed orbits arising in the context of the closed-orbit
theory of atomic photoionization. To improve these results, Sieber and
Schomerus \cite{Sie96,Sch97a,Sie98} recently derived uniform approximations
which interpolate smoothly between Gutzwiller's isolated-orbits
contributions on either side of the bifurcation. Their formulas are
applicable to all kinds of period-$m$-tupling bifurcation generic to
Hamiltonian systems with two degrees of freedom.

A closer inspection of bifurcation scenarios encountered in practical
applications of semiclassical quantization reveals, however, that the
uniform approximations applicable to generic codimension-one
bifurcations need to be extended to also include bifurcations of
higher codimension. Although these non-generic bifurcations cannot
directly be observed if only a single control parameter is varied,
they can nevertheless produce clear signatures in semiclassical
spectra because in their neighborhood two codimension-one bifurcations
approach each other, so that all periodic orbits involved in any of
the subsequent bifurcations have to be treated collectively. Examples
of this situation were studied by Schomerus and Haake
\cite{Sch97b,Sch97c} as well as by Main and Wunner
\cite{Mai97a,Mai98a}, who applied techniques of catastrophe theory to
achieve a collective treatment of complicated bifurcation scenarios.

All uniform approximations discussed so far in the literature require
the inclusion of complex ``ghost orbits''. At a bifurcation point, new
real periodic orbits are born. If, in the energy range where the real
orbits do not exist, the search for periodic orbits is extended to the
complexified phase space, the orbits about to be born can be found to
possess complex predecessors -- ghost orbits. As was first shown by
Ku\'s et.~al.~\cite{Kus93}, some of these ghost orbits, whose
contributions become exponentially small in the limit of $\hbar \to
0$, have to be included in Gutzwiller's trace formula. In addition,
the construction of uniform approximations requires complete
information about the bifurcation scenario, including ghost orbits.

All bifurcation scenarios discussed in the physics literature so far
involved bifurcations of real orbits only. However, there is no reason
why ghost orbits should not themselves undergo bifurcations in their
process of turning real. It is the purpose of this work to demonstrate
that ghost orbit bifurcations do indeed occur and have a pronounced
effect on semiclassical spectra if they arise as part of a bifurcation
scenario of higher codimension. To this end, we present an example
taken from the Diamagnetic Kepler Problem. The example we chose
appears to be simple: We discuss the period-quadrupling of the balloon
orbit, which is one of the shortest periodic orbits in the Diamagnetic
Kepler Problem. However, even this simple case turns out to require
the inclusion of a ghost orbit bifurcation.

To cope with this new situation, we have to develop a technique which
enables us to deal with the occurence of ghost orbits. It turns out
that normal form theory allows for a description of real and ghost
orbit bifurcations on an equal footing. Consequently, ghost orbit
bifurcations are found to contribute to uniform approximations in much
the same way as bifurcations of real orbits do, provided that they
occur in connection to bifurcations of real orbits as part of a
bifurcation scenario of higher codimension. Therefore, we will arive
at the conclusion that in generic Hamiltonian systems with mixed
regular-chaotic dynamics the occurence of ghost orbit bifurcations
will not be a very exotic, but rather quite a common phenomenon.

The organization of this paper is as follows: In section 2, we briefly
summarize the derivation of Gutzwiller's trace formula, which forms the
basis of semiclassical theories of the density of states. Section 3
presents the bifurcation scenario of the example chosen. In section 4, we
discuss normal form theory and show that it allows for an analytic
description of the example bifurcation scenario. Section 5 then contains
the uniform approximation pertinent to the example scenario. It is
evaluated in two different degrees of approximation, one of which
asymptotically yields perfect agreement with the results of Gutzwiller's
trace formula. A more concise presentation of our main results can be found
in  \cite{Bar99}.

\section{Gutzwiller's trace formula}
\label{SpurfSec}

Gutzwiller's trace formula offers a way to calculate a semiclassical
approximation to the quantum mechanical density of states
\begin{equation}\label{dDef}
  d(E) = \sum_j \delta(E-E_j) \;,
\end{equation}
where the sum extends over all quantum eigenenergies $E_j$ of the
system under study. In quantum mechanics, the density of states is given by
\begin{equation}\label{Spur}
  d(E) = \sum_j\delta(E-E_j) = -\frac{1}{\pi}\,\Im\Tr G \;,
\end{equation}
where
\begin{equation*}
  \Tr G := \int d^nx\,G(\vec{x}\vec{x},E) \;,
\end{equation*}
and the Green's function $G(\vec x'\vec x, E)$ is the configuration space
representation of the resolvent operator
\begin{equation*}
  \hat G(E) = (E-\hat H)^{-1} \;.
\end{equation*}
On the other hand, it is connected to the time-domain propagator $K(\vec
x't', \vec xt)$ by a Fourier transform
\begin{equation}\label{Green}
  G(\vec x'\vec x, E) = \frac{1}{i\hbar} \int_0^\infty dt\,
    K(\vec x't', \vec xt) \exp\left\{i\frac{Et}{\hbar}\right\} \;.
\end{equation}
Thus, if Feynman's path integral representation of the propagator
\begin{equation}
  K(\vec{x}'t',\vec{x}t) = \int {\cal D}\left(\vec{x}(t)\right) \,
    \exp\left\{\frac{i}{\hbar}\int_{t}^{t'}dt\,
      L\left(\vec{x}(t),\dot{\vec{x}}(t),t\right)\right\}
\end{equation}
and the Fourier integral are approximately evaluated by the method of
stationary phase, one obtains the semiclassical Green's function
(\cite{Gut67}, see also \cite{Schu81})
\begin{equation}\label{sclGreen}
  G_{scl}(\vec{x}'\vec{x},E) = \frac{2\pi}{(2\pi i\hbar)^{(n+1)/2}}
    \sum_{\text{class.traj.}}\sqrt{|D|}
    \exp\left\{i\left(\frac{S}{\hbar}-\mu\frac{\pi}{2}\right)\right\}
  \;.
\end{equation}
Here, the sum extends over all classical trajectories connecting $\vec x$
to $\vec x'$ at energy $E$,
\begin{equation}\label{SDef}
  S(\vec{x}'\vec{x},E) = \int \vec{p}\cdot d\vec{x}
\end{equation}
denotes the classical action along the trajectory,
\begin{equation*}
  D  = \det\begin{pmatrix}
         \ppabl{S}{\vec{x}'}{\vec{x}}&\ppabl{S}{\vec{x}'}{E}\\[1ex]
         \ppabl{S}{E}{\vec{x}} & \PPabl{S}{E}
       \end{pmatrix} \;,
\end{equation*}
and the integer $\mu$ counts the number of caustics the trajectory touches.

To find the density of states, one has to calculate the trace of the
semiclassical Green's function. To this end, one calculates the limit of
$G_{scl}$ for $\vec x'\to\vec x$ and then integrate over $\vec x$. If $\vec
x'$ is very close to $\vec x$, there always exists a direct path connecting
$\vec x$ to $\vec x'$. In addition, there are usually indirect paths which
leave the neighborhood of their starting point before returning there. The
contribution of the direct path can be shown to yield \emph{Weyl's density
  of states}
\begin{equation}\label{Weyl}\begin{split}
  \bar{d}(E) &= \frac{1}{(2\pi\hbar)^n} \int d^nx\,d^np\,
    \delta\left(E-\frac{\vec{p}^2}{2m}-V(\vec{x})\right) \\
  &= \frac{1}{(2\pi\hbar)^n} \frac{d}{dE}
    \underbrace{\int d^nx\,d^np\,
        \Theta\left(E-\frac{\vec{p}^2}{2m}-V(\vec{x})\right)\;,}
      _{\text{classically accessible volume of phase space}}
\end{split}\end{equation}
where $\Theta$ denotes Heaviside's step function. This result reproduces
the well-known fact from statistical mechanics that on the average there is
one quantum state per phase space volume of $(2\pi\hbar)^n$. The
contributions of indirect paths then superimpose system-specific
modulations on this general average value.

Due to the stationary-phase condition, only periodic orbits contribute to
the semiclassical density of states. To determine the contribution of a
single periodic orbit, one introduces a coordinate system with one
coordinate running along the periodic orbit and all other coordinates
perpendicular to it. Assuming all periodic orbits to be isolated in phase
space, one can then evaluate the trace by the method of stationary phase
and obtains \emph{Gutzwiller's trace formula} for the system-specific
modulations of the density of states
\begin{equation}\label{Spurf}
  d'(E) = \frac{1}{\pi\hbar}\,\Re \sum_{\text{po}}
    \frac{T_0}{\sqrt{\left|\det(M-I)\right|}}
    \exp\left\{\frac{i}{\hbar}S -i\frac{\pi}{2}\nu\right\} \;.
\end{equation}
Here, the sum runs over all periodic orbits at energy $E$, $S$ denotes the
action of the orbit, $T_0$ its primitive period, $M$ its monodromy matrix,
which describes the stability of the orbit, and $\nu$ its Maslov
index, which reflects the topology of nearby orbits. In the derivation, the
primitive period $T_0$ can be seen to arise from the integration along the
orbit, whereas the occurrence of the monodromy matrix is due to the
integrations over the transverse coordinates.

Gutzwiller's trace formula expresses the quantum density of states in terms
of purely classical data. It fails, however, if the periodic orbits of the
classical system cannot be regarded as isolated, as is the case, e.g.,
close to a bifurcation. There, the failure of the trace formula manifests
itself in a divergence of the isolated-orbits contributions in
(\ref{Spurf}): If an orbit undergoes a bifurcation, the determinant of
$M-I$ vanishes. In recent years, the problem of calculating the joint
contribution of bifurcating orbits to the density of states was addressed
by various authors, whose works were briefly reviewed in the introduction.
It is the purpose of the present paper to present normal form theory as a
technique which allows one to achieve a \emph{collective} description of
bifurcating orbits and to show its applicability to a complicated
bifurcation scenario. In particular, we shall demonstrate that bifurcations
of ghost orbits need to be included in the description of classical
bifurcation scenarios because they can exert a marked influence on
semiclassical spectra, and that classical normal form theory can be
extended so as to meet this requirement. However, before we come to deal
with the construction of uniform approximations in sections \ref{NFBifSec}
and \ref{UnifLsgSec}, we shall give a description of our example system,
the Diamagnetic Kepler Problem, and the bifurcation scenario we are going
to study.

\section{The Diamagnetic Kepler Problem}
\label{BifSec}

\subsection{The Hamiltonian}
As a prototype example of a system which undergoes a transition to chaos,
we shall investigate the hydrogen atom in a homogeneous external magnetic
field, which is reviewed, e.g., in \cite{Fri89,Has89,Wat93}. We assume the
nucleus fixed and regard the electron as a structureless point charge
moving under the combined influences of the electrostatic Coulomb force and
the Lorentz force.  Throughout this paper, we shall use atomic units, let
the magnetic field point along the $z$-direction and denote its strength by
$\gamma = B/B_0$, where
$B_0=m^2e^3(4\pi\varepsilon_0)^{-2}\hbar^{-3}=2.3505\cdot10^5 \rm T$ is the
atomic unit of the magnetic field strength. The Hamiltonian then reads
\begin{equation}\label{HamFkt}
 H = \frac{\vec{p}^{\,2}}{2} +\frac{\gamma}{2}L_z 
      +\frac{\gamma^2}{8}\left(x^2+y^2\right) -\frac{1}{r} \;,
\end{equation}
where $r=|\vec{x}|$ and
$L_z=\vec{e}_z\cdot\left(\vec{x}\times\vec{p}\right)$ denotes the
$z$-component of the angular momentum. In the following, we will restrict
ourselves to the case $L_z = 0$. As a consequence, the angular coordinate
$\varphi$ measuring rotation around the field axis becomes ignorable, so
that we are effectively dealing with a two-degree-of-freedom system.

The energy $E=H$ is a constant of motion. Thus, the dynamics depends on
both the energy and the magnetic field strength as control parameters. This
situation can be simplified, however, if one exploits the scaling
properties of the Hamiltonian. If classical quantities are scaled according
to
\begin{equation}\label{WirkSkal}
  \begin{array}{r@{}l@{}l@{\qquad\qquad}r@{}l@{}l}
    \vec{x}&\mapsto\tilde{\vec{x}}&=\gamma^{2/3}\vec{x}\;, &
    \vec{p}&\mapsto\tilde{\vec{p}}&=\gamma^{-1/3}\vec{p}\;, \\
    H&\mapsto\tilde{H}&=\gamma^{-2/3}H\;, &
    t&\mapsto\tilde{t}&=\gamma t\;, \\
    S&\mapsto\tilde{S}&=\gamma^{1/3}S\;,
\end{array}\end{equation}
one obtains the scaled Hamiltonian
\begin{equation}\begin{split}
  \tilde{H} &= \frac{\tilde{\vec{p}}^2}{2}
    +\frac{1}{8}\left(\tilde{x}^2+\tilde{y}^2\right)
    -\frac{1}{\tilde{r}}\\
  &= \tilde{E} = \gamma^{-2/3}E \;.
\end{split}\end{equation}
The scaled dynamics depends on the scaled energy as its only control
parameter.

The equations of motion following from this Hamiltonian are difficult to
handle numerically due to the Coulomb singularity at $\tilde r = 0$. To
overcome this problem, one introduces \emph{semiparabolical coordinates}
\begin{equation}\label{SPKoord}
  \mu^2=\tilde{r}-\tilde{z}\;,\qquad\qquad
  \nu^2=\tilde{r}+\tilde{z}
\end{equation}
and a new orbital parameter $\tau$ defined by
\begin{equation*}
  dt = 2r \,d\tau = \left(\mu^2+\nu^2\right) \,d\tau \;.
\end{equation*}
These transformations lead to the final form of the Hamiltonian
\begin{equation}\label{regHamFkt}
  \cH = \frac{p_\mu^2+p_\nu^2}{2}-\tilde{E}\left(\mu^2+\nu^2\right)
       +\frac{1}{8}\mu^2\nu^2\left(\mu^2+\nu^2\right) \equiv 2 \;.
\end{equation}
In this form, the scaled energy $\tilde E$ plays the r\^ole of an external
parameter, whereas the value of the Hamiltonian is fixed: It has to be
chosen equal to 2. The equations of motion following from this Hamiltonian
do not contain singularities any more so that they can easily be integrated
numerically.  

Note that the definition (\ref{SPKoord}) determines the
semiparabolical coordinates up to a choice of sign only. Thus, orbits which
are mirror images of each other with respect to a reflection at the $\mu$-
or $\nu$-axes have to be identified. Furthermore, if we follow a periodic
orbit until it closes in $(\mu,\nu)$-coordinates, this may correspond to
more than one period in the original configuration space. This has to be
kept in mind when interpreting plots of periodic orbits in semiparabolical
coordinates.

As a substantial extension of the classical description of the Diamagnetic
Kepler Problem we complexify the classical phase space by allowing
coordinates and momenta to assume complex values. As the Hamiltonian
(\ref{regHamFkt}) is holomorphic, we can at the same time regard the phase
space trajectories as functions of complex times $\tau$. To numerically
calculate the solution of the equations of motion at a given time $\tau_f$,
we integrate the equations of motion along a path connecting the origin of
the complex $\tau$-plane to the desired endpoint $\tau_f$. By Cauchy's
integral theorem, the result does not depend on the path chosen so that we
can safely choose to integrate along a straight line from $0$ to $\tau_f$.
This extension allows us to look for ghost orbit predecessors of real
periodic orbits born in a bifurcation. In general, their orbital parameters
$S$, $T$ and the monodromy matrix $M$ will be complex. We calculate them
along with the numerical integration of the equations of motion from
\begin{equation*}
  T = \int_0^{\tau_f} (\mu^2+\nu^2)\,d\tau \;,\qquad
  S = \int_0^{\tau_f} (p_\mu^2+p_\nu^2)\,d\tau \;.
\end{equation*}

\subsection{The bifurcation scenario}
The Diamagnetic Kepler Problem described by the Hamiltonian
(\ref{regHamFkt}) exhibits a transition between regular dynamics at
strongly negative scaled energies $\tilde E\to -\infty$ and chaotic
dynamics at $\tilde E\approx 0$ and above (for details see, e.g.,
\cite{Has89}). Correspondingly, there are only three different
periodic orbits at very low scaled energy, whereas the number of
periodic orbits increases exponentially as $\tilde E\nearrow 0$.

\begin{figure}
  \vspace{18.0cm}
  \includegraphics{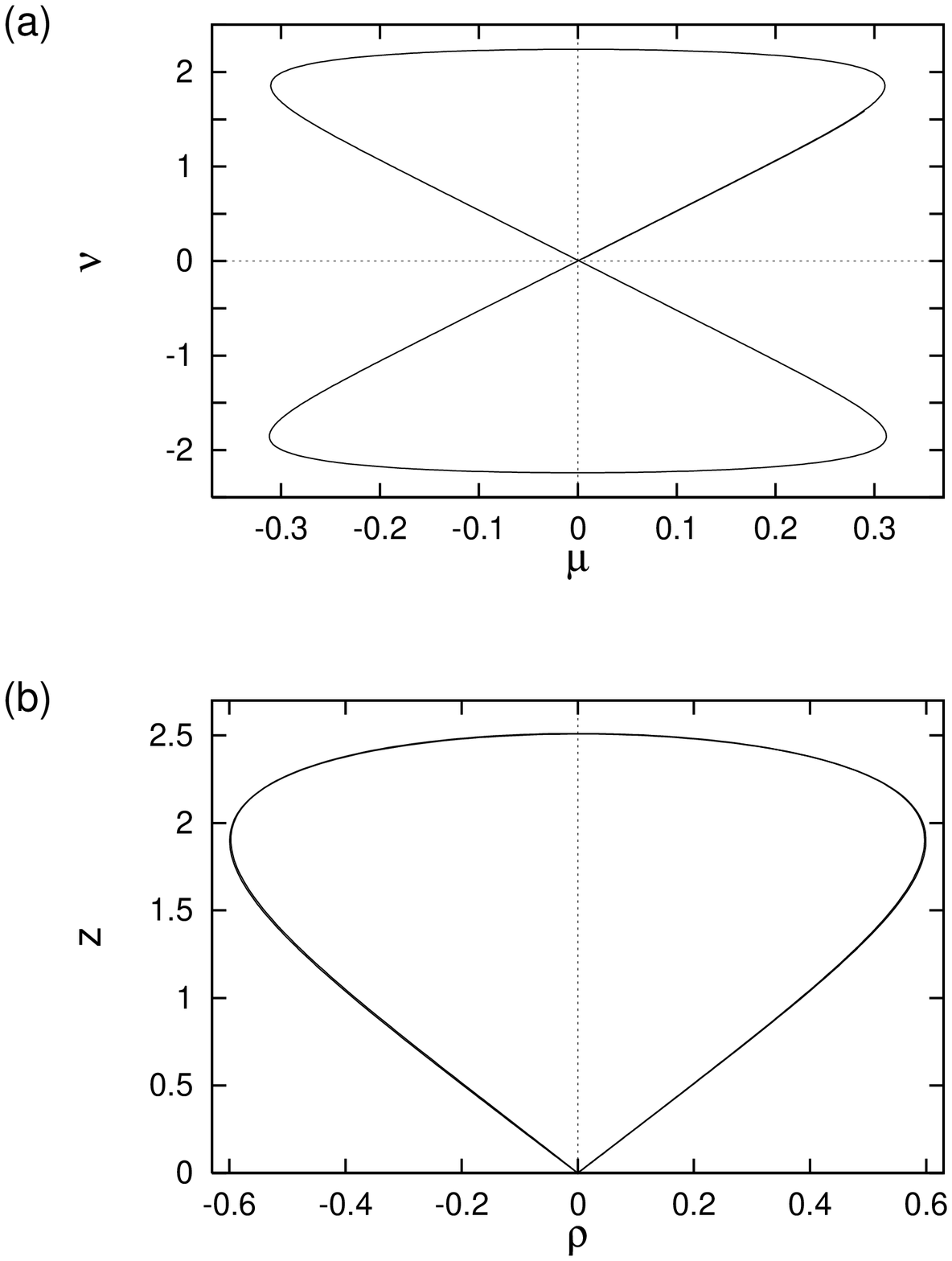}
  \caption{\label{BalAbb}
    The balloon orbit (a) in semiparabolical and (b) in cylindrical
    coordinates at a scaled energy of $\tilde E=-0.34$.}
\end{figure}

\begin{figure}
  \wlabelgfx{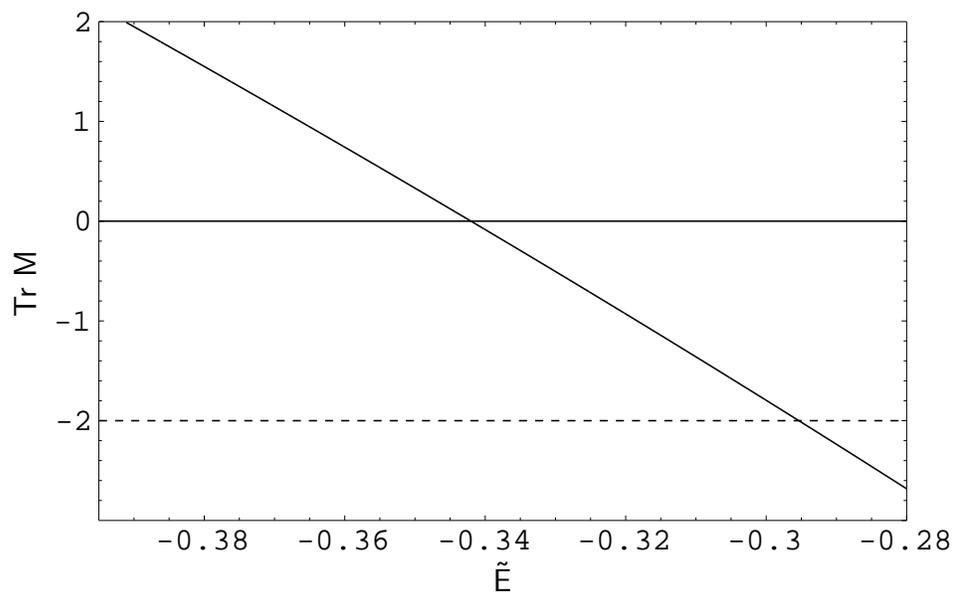}
              {\mbox{\large $\mathsf{\tilde E}$}}
              {\mbox{\large\sffamily Tr M}}{12cm}                                 \caption{\label{SpurAbb}
    The trace of the monodromy matrix of the balloon orbit. The zero
    at $\tilde E_c = -0.342025$ indicates the occurence of the
    period-quadrupling bifurcation.}
\end{figure}

At any fixed scaled energy, there is a periodic orbit parallel to the
magnetic field. It is purely Coulombic since a motion parallel to the
magnetic field does not cause a Lorentz force. This orbit is stable at low
negative scaled energies; as $\tilde E\nearrow 0$, however, it turns
unstable and stable again infinitely often \cite{Win87}. For the first
time, instability occurs at $\tilde E = -0.391$. In this bifurcation, a
stable and an unstable periodic orbit are born. The stable orbit is known
as the \emph{balloon orbit}. It is depicted in figure \ref{BalAbb}. As
shown in bifurcation theory, the stability of a periodic orbit is
determinded by the trace of its monodromy matrix.  For the balloon orbit,
the trace is shown in figure \ref{SpurAbb}. It equals 2 when the orbit is
born. As the scaled energy increases, the trace decreases monotonically.
The orbit turns unstable at $\tilde E=-0.291$, where the trace equals $-2$.
In between, all kinds of period-$m$-tupling bifurcations occur. In this
work, we shall discuss the period-quadrupling bifurcation which arises at
the zero of the trace at $\tilde E_c=-0.342025$.

\begin{figure}
  \vspace{18.0cm}
  \includegraphics{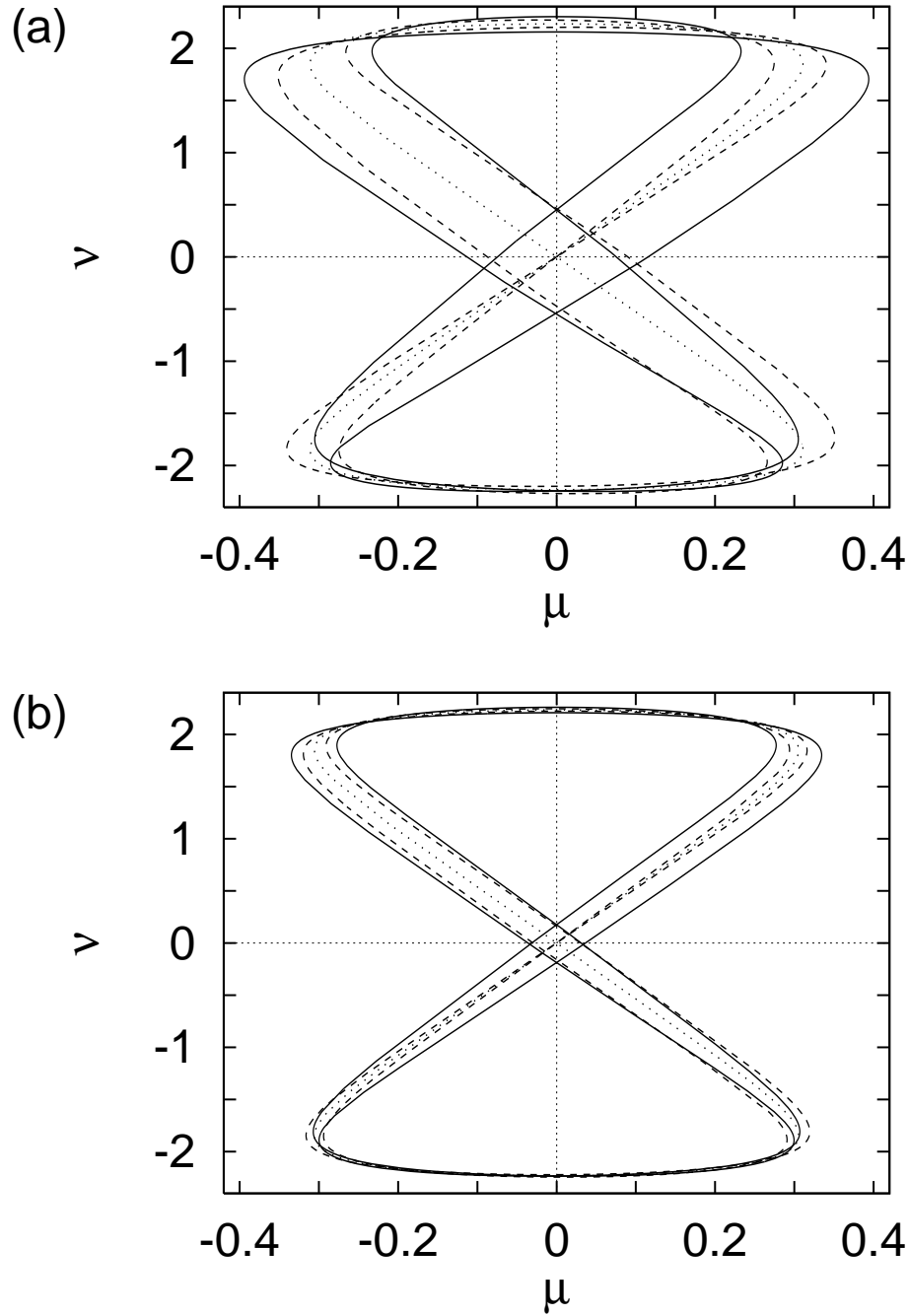}
  \caption{\label{reellAbb}
    Real satellite orbits involved in the period-quadrupling
    bifurcation of the balloon orbit at scaled energies of (a) $\tilde
    E = -0.34$ and (b) $\tilde E = -0.3418$. Solid curve: stable
    satellite, dashed curve: unstable satellite; for comparison:
    dotted curve: balloon orbit.}
\end{figure}

\begin{figure}
  \vspace{18.0cm}
  \includegraphics{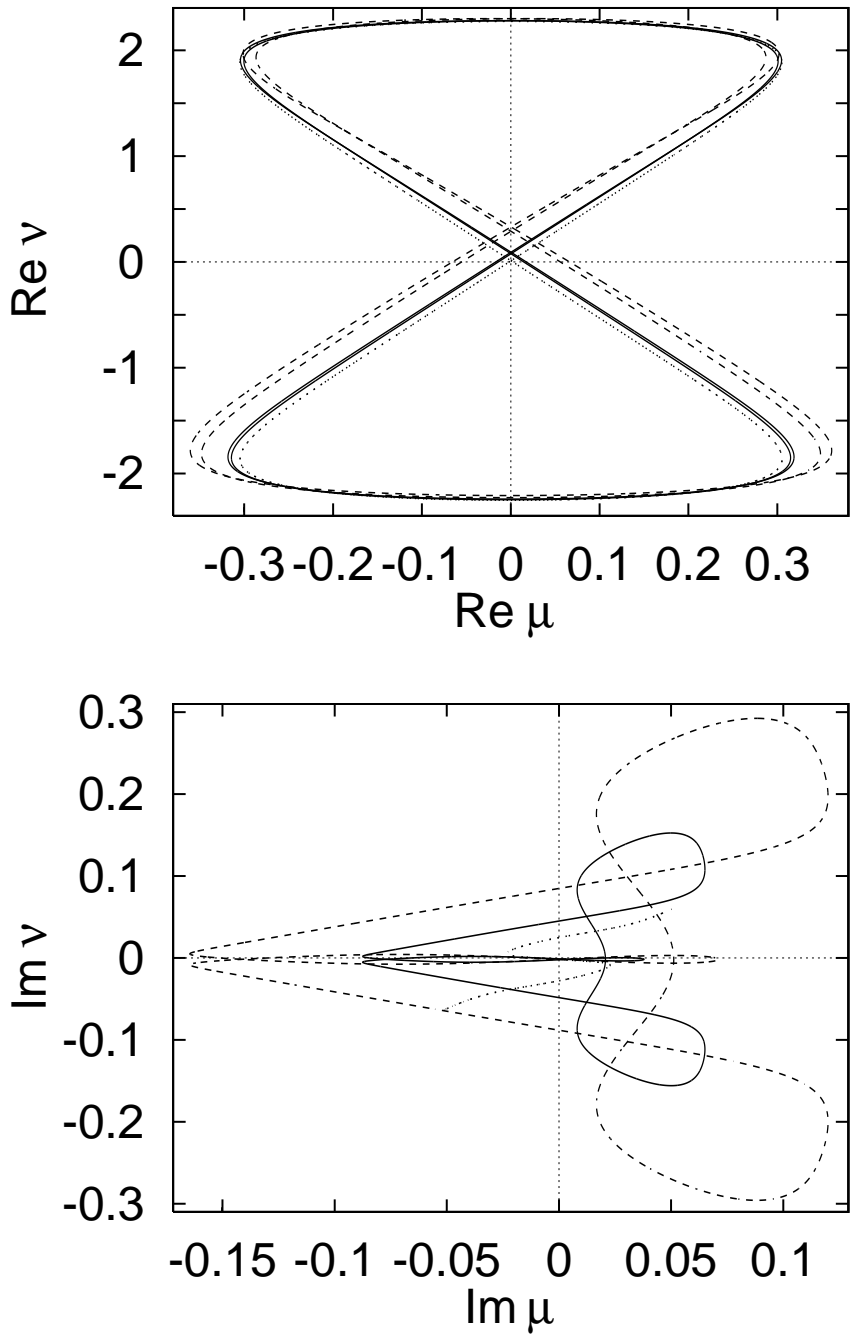}
  \caption{\label{Kplx343}
    Ghost orbits at scaled energy of $\tilde{E} = -0.343$. Solid and
    dotted curves: stable and unstable ghost satellite orbits created
    in the period-quadrupling of the balloon orbit at $\tilde
    E_c=-0.342025$.  Dashed curve: additional ghost orbit created in
    the ghost bifurcation at $\tilde E_c' = -0.343605$.}
\end{figure}

For $\tilde E>\tilde E_c$, two real satellite orbits of quadruple period
exist. They are depicted in figure \ref{reellAbb} at two different values
of the scaled energy. The solid and dashed curves in the plots represent
the stable and unstable satellite orbits, respectively. In both cases, the
balloon orbit is shown for comparison as a dotted curve. The satellites can
clearly be seen to approach the balloon orbit as $\tilde E\searrow\tilde
E_c$. At $\tilde E_c$, they collide with the balloon orbit and disappear.
Below $\tilde E_c$, a stable and an unstable ghost satellite exist instead.
They are presented as the solid and dotted curves in figure \ref{Kplx343}.
Note that the imaginary parts are small compared to the real parts because
the bifurcation where the imaginary parts vanish is close. As the
Hamiltonian (\ref{regHamFkt}) is real, the complex conjugate of any orbit
is again a solution of the equations of motion. In this case, however, the
ghost satellites coincide with their complex conjugates, so that the total
number of orbits is conserved in the bifurcation.  This behavior can be
understood in terms of normal form theory (see section \ref{GenBifSec}).

\begin{figure}
  \vspace{18.0cm}
  \includegraphics{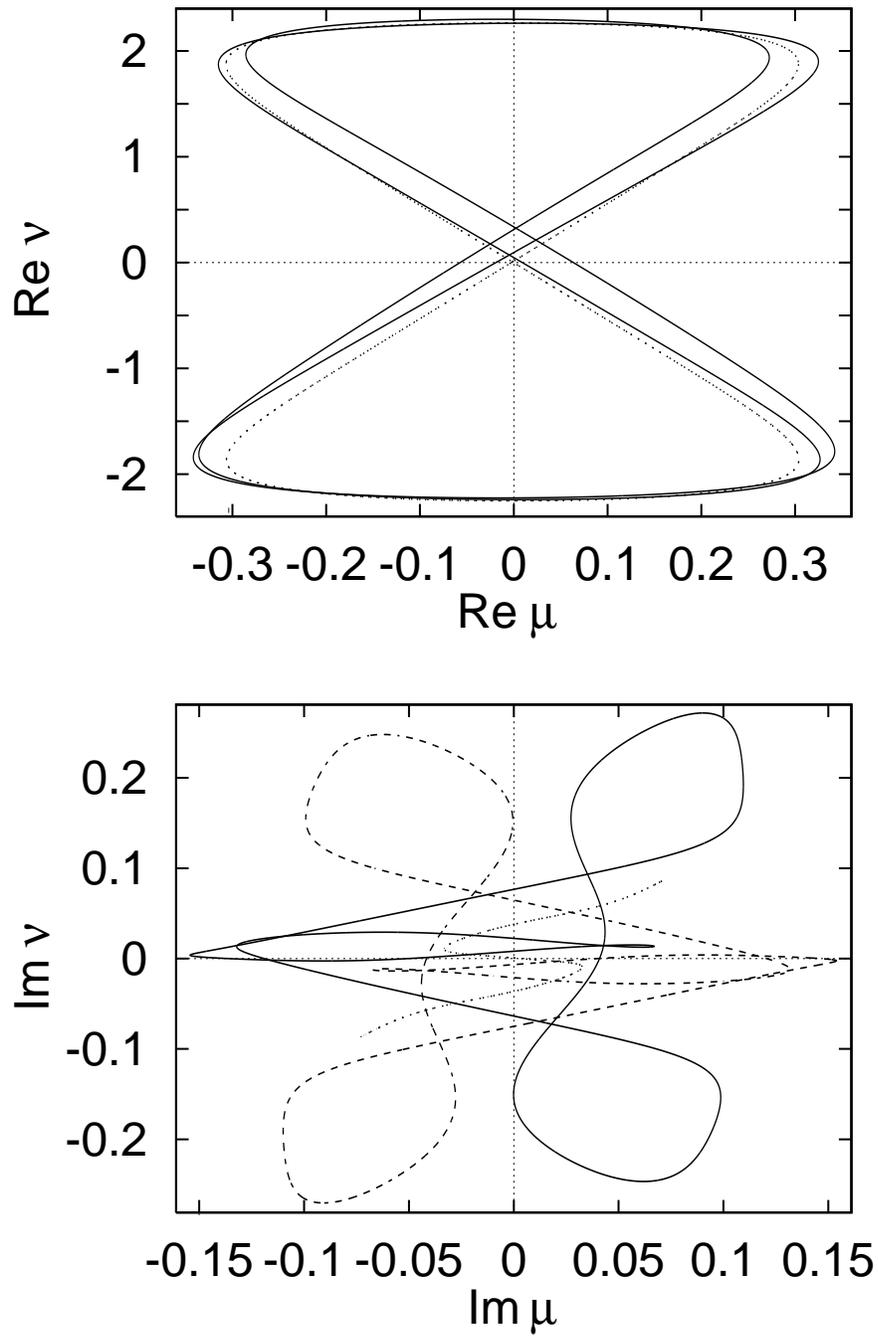}
  \caption{\label{Kplx344}
    Ghost orbits at scaled energy of $\tilde{E} = -0.344$. Solid and
    dashed curves: Asymmetric ghost orbits created in the ghost
    bifurcation at $\tilde E_c' = -0.343605$ (real parts
    coincide). Dotted curve: Unstable ghost satellite orbit created in
    the period quadrupling of the balloon orbit at $\tilde
    E_c=-0.342025$.}
\end{figure}

The orbits described so far form a generic kind of period-quadrupling
bifurcation as described by Meyer \cite{Mey70} and dealt with in the
context of semiclassical quantization by Sieber and Schomerus \cite{Sie98}.
In our case, however, this description of the bifurcation scenario is not
yet complete because there exists an additional periodic ghost orbit at
scaled energies around $\tilde E_c$. Its shape is shown as a dashed curve
in figure \ref{Kplx343}. It is very similar to the stable ghost satellite
originating in the period-quadrupling, and indeed, when following the ghost
orbits to lower energies, we find another bifurcation at $\tilde E_c' =
-0.343605$, i.e., only slightly below the bifurcation point $\tilde
E_c=-0.342025$ of the period-quadrupling. At $\tilde E_c'$, the additional
ghost orbit collides with the stable ghost satellite, and these two orbits
turn into a pair of complex conjugate ghost orbits. Their shapes are
presented at a scaled energy of $\tilde E = -0.344$ as the solid and dashed
curves in figure \ref{Kplx344}. From the imaginary parts, the loss of
conjugation symmetry can clearly be seen if the symmetries of the
semiparabolical coordinate system as described above are taken into
account. The dotted curve in figure \ref{Kplx344} represents the unstable
ghost satellite which was already present at $\tilde E>\tilde E_c'$. It
does not undergo any further bifurcations.

Note that the second bifurcation at $\tilde E = \tilde E_c'$ involves
ghost orbits only. This kind of bifurcation has not yet been described
in the literature so far. In particular, Meyer's classification of
codimension-one bifurcations in generic Hamiltonian systems covers
bifurcations of real orbits only and does not include bifurcations of
ghost orbits. Consequently, the influence of ghost orbit bifurcations
on semiclassical spectra has never been investigated so far. Due to
the existence of this bifurcation, however, the results by Sieber and
Schomerus \cite{Sie98} concerning generic period-quadrupling
bifurcations cannot be applied to the complicated bifurcation scenario
described here. As in cases dealt with before by Main and Wunner
\cite{Mai97a,Mai98a} as well as by Schomerus and Haake
\cite{Sch97a,Sch97b}, who discussed the semiclassical treatment of two
neighboring bifurcations of real orbits, the closeness of the two
bifurcations requires the construction of a uniform approximation
taking into account all orbits involved in either bifurcation
collectively. Thus, the ghost orbit bifurcation at $\tilde E_c'$ turns
out to contribute to the semiclassical spectrum in much the same way
as a bifurcation of real orbits does, as long as we do not go to the
extreme semiclassical domain where the two bifurcations can be
regarded as isolated and ghost orbit contributions vanish altogether.

\begin{figure}
  \vspace{18.0cm}
  \includegraphics{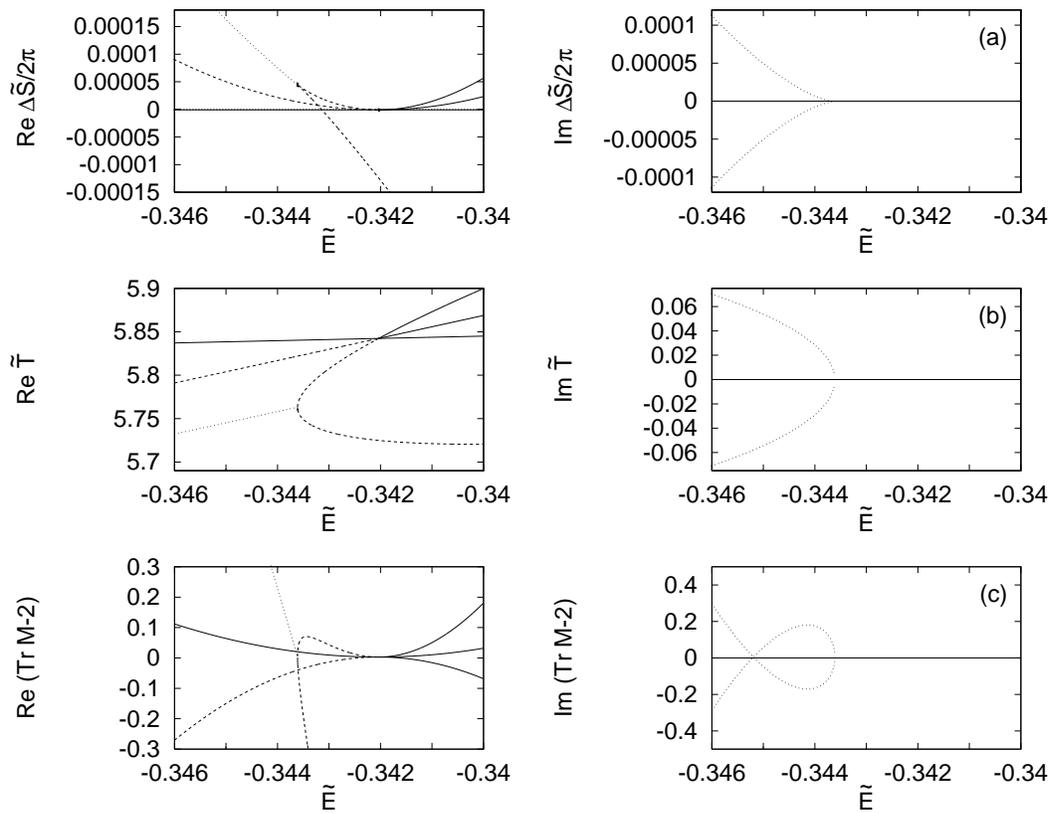}
  \caption{\label{DataFig}
    Actions, orbital periods and traces of the monodromy matrices of
    the orbits involved in the bifurcation scenario as functions of
    the scaled energy $\tilde E = \gamma^{-2/3}E$. Solid curves: real
    orbits, dashed curves: ghost orbits symmetric with respect to
    complex conjugation, dotted curves: asymmetric ghost orbits.}
\end{figure}

To construct a uniform approximation, we need to know the periodic orbit
parameters of all orbits involved in the bifurcations. The parameters were
calculated numerically and are displayed in figure \ref{DataFig} as
functions of the scaled energy. Part (a) of this figure presents the
actions of the orbits. To exhibit the sequence of bifurcations more
clearly, the action of four repetitions of the central balloon orbit was
chosen as a reference level $(\Delta S=0)$. Around $\tilde E_c$, we
recognize two almost parabolic curves which indicate the actions of the
stable (upper curve) and unstable (lower curve) satellite orbits. At
$\tilde E_c$, the curves change from solid to dashed as the satellite
orbits become complex. Another dashed line represents the action of the
additional ghost orbit, which can clearly be seen not to collide with the
balloon orbit at $\tilde E_c$. Whereas the unstable ghost satellite does
not undergo any further bifurcations, the curves representing the stable
and the additional ghost orbits can be seen to join at $\tilde
E_c'=-0.343605$. Below $\tilde E_c'$, a dotted curve indicates the presence
of a pair of complex conjugate ghosts. The imaginary parts of their actions
are different from zero and have opposite signs, whereas above $\tilde
E_c'$, all ghost orbits coincide with their complex conjugates so that
their actions are real.

Analogously, part (b) of figure \ref{DataFig} presents the orbital periods.
In this case, no differences were taken, so that the period of the fourth
repetition of the balloon orbit, which is always real, appears in the
figure as a nearly horizontal line at $T\approx 5.84$. The other orbits can
be identified with the help of the bifurcations they undergo, similar to
the discussion of the actions given above. Finally, figure \ref{DataFig}(c)
shows the traces of the monodromy matrices minus two. For Hamiltonian
systems with two degrees of freedom, these quantities agree with $\det
(M-I)$. At $\tilde E_c$ and $\tilde E_c'$, they can be seen to vanish for
the bifurcating orbits, thus causing the periodic orbit amplitudes
(\ref{Spurf}) to diverge at the bifurcation points.

\section{Normal form theory and bifurcations}
\label{NFBifSec}

\subsection{Birkhoff-Gustavson normal form}
As we have seen, Gutzwiller's trace formula (\ref{Spurf}) fails close to
bifurcations, when periodic orbits of the classical system cannot be
regarded as isolated. To overcome this difficulty, we need a technique
which allows us to describe the structure of the classical phase space
close to a bifurcating orbit. This can be done with the help of normal form
theory. A detailed description of this technique can be found in
\cite[Appendix 7]{Arn88a} or \cite[Sections 2.5 and 4.2]{Alm88}. Here, we
will present the normal form transformations for systems with two degrees
of freedom only, although the general scheme is the same for higher
dimensional cases. As only stable periodic orbits can undergo bifurcations,
we will restrict ourselves to this case.

\sloppypar
As a first step, we introduce a special canonical coordinate system
$(\vartheta, p_\vartheta, q, p)$ in a neighborhood of a periodic
orbit, which has the following properties (concerning the existence of
such a coordinate system, see \cite[Chapter 7.4, Proposition
1]{Arn88b}): 
\begin{itemize}
\item
  $\vartheta$ is measured along the periodic orbit and $q$
  perpedicular to it, so that phase space points lying on the periodic
  orbit are characterized by $q=p=0$.
\item
  $\vartheta$ assumes values between 0 and $2\pi$, and along the
  periodic orbit we have, up to a constant,
  \begin{equation*}
    \vartheta = \frac{2\pi}{T}t\;,
  \end{equation*}
  where $T$ denotes the orbital period.
\item
  If we choose an initial condition in the neighborhood and
  $0\le\vartheta<T$, the function $\vartheta(t)$ is invertible.
\end{itemize}
According to the last condition, we can regard $p$ and $q$ as
functions of $\vartheta$ instead of $t$.

The classical dynamics of a mechanical system is given by Hamilton's
variational principle, which states that a classical trajectory with fixed
initial and final coordinates $\vec{q}(t_1)$ and $\vec q(t_2)$ satisfies
\begin{equation}\label{HamPrinz}
  \delta\,\int_{\left(\vec{q}(t_1),t_1\right)}
              ^{\left(\vec{q}(t_2),t_2\right)}
           \vec{p}\cdot d\vec{q} - H\,dt
  = 0 \;.
\end{equation}
If we restrict ourselves to considering the energy surface given by a fixed
energy $E$, we can transform the integral as follows:
\begin{equation}\begin{split}
    \int \vec{p}\cdot d\vec{q} - H \, dt 
  &=\int p \, dq + p_\vartheta \,d\vartheta - H\,dt  \\
  &=\int p \, dq - (-p_\vartheta) \,d\vartheta \quad - E (t_2-t_1) \;.
\end{split}\end{equation}
The last term in this expression is a constant which does not
contribute to the variation of the integral, so that actual orbits of
the system satisfy
\begin{equation}
  \delta\!\int p \, dq - (-p_\vartheta) \,d\vartheta
  \;=\;\delta\!\int \vec{p}\cdot d\vec{q} - H \, dt
  \;=\; 0 \;.
\end{equation}
Thus, the dependence of $p$ and $q$ on the new parameter $\vartheta$
is given by Hamilton's equations of motion, where $-p_\vartheta$ plays
the r\^ole of the Hamiltonian. It has to be determined as a function
of the new phase space coordinates $p, q$, the "time" $\vartheta$ and
the energy $E$, which occurs as a parameter, from the equation
\begin{equation*}
  H\left(\vartheta,p_\vartheta,q,p\right) = E
\end{equation*}
From our choice of the coordinate system, $p_\vartheta$ is periodic in
$\vartheta$ with a period of $2\pi$.

We have now reduced the dynamics of a two-degrees-of-freedom
autonomous Hamiltonian system to that of a single-degree-of-freedom
system, which is, however, no longer autonomous, but periodically
time-dependent. With regard to the original system, we can view the
motion perpendicular to the periodic orbit as being periodically
driven by the motion along the orbit. Henceforth, we shall denote the
Hamiltonian of the reduced system by $\Phi$, coordinate and momentum
by $q$ and $p$, respectively, and the time by $\vartheta$.

The point $p=q=0$ corresponds to the periodic orbit of the original
system and therefore constitutes a stable equilibrium position of the
reduced system so that a Taylor series expansion of the Hamiltonian
around this point does not have linear terms. By a suitable
time-dependent canonical transformation, the quadratic term can be
made time-independent, see \cite{Sad96}. We expand the Hamiltonian
in a Taylor series in $p$ and $q$ and in a Fourier series in
$\vartheta$:
\begin{equation}
  \Phi(p,q,\vartheta) = \frac{\omega}{2}(p^2+q^2) + 
  \sum_{\alpha+\beta=3}^\infty \sum_{l=-\infty}^\infty
    \Phi_{\alpha\beta l} p^\alpha q^\beta \exp(il\vartheta) \;.
\end{equation}
To go on, we introduce complex coordinates 
$$ z = p+iq \qquad\qquad z^\ast = p-iq \;.$$
This transformation is canonical with multiplier $-2i$, so that we
have to go over to a new Hamiltonian
\begin{equation}\begin{split}
  \label{NFAnsatz}
  \phi &= -2 i \Phi \\
   &= -i\omega z z^\ast + \sum_{\alpha+\beta=3}^\infty
      \sum_{l=-\infty}^\infty \phi_{\alpha\beta l}z^\alpha z^{\ast\beta}
          \exp(il\vartheta)
\end{split}\end{equation}

Birkhoff \cite[Chapter 3]{Bir27} and Gustavson \cite{Gus66} developed
a technique which allows us to systematically eliminate low-order
terms from this expansion by a sequence of canonical
transformations. To eliminate terms of order $\alpha+\beta=k$, we
employ the transformation given by the generating function
\begin{equation}
  {\cal F}_2(Z^\ast,z,\vartheta) = -z Z^\ast -
    \sum_{\alpha+\beta=k} \sum_{l=-\infty}^{\infty} 
      {\cal F}_{\alpha\beta l} z^\alpha Z^{\ast\beta} \exp(il\vartheta)
  \label{ordkErz}
\end{equation}
with arbitrary expansion coefficients ${\cal F}_{\alpha\beta l}$,
so that the transformation reads
\begin{equation}\label{ordkTrafo}\begin{aligned}
  Z = -\frac{\partial{\cal F}_2}{\partial Z^\ast}
   &= z \left( 1+\sum\beta{\cal F}_{\alpha\beta l} z^{\alpha -1}
        Z^{\ast\beta-1}\exp(il\vartheta) \right) \;,\\
  z^\ast = -\frac{\partial{\cal F}_2}{\partial z}
   &= Z^\ast \left(1+\sum \alpha {\cal F}_{\alpha\beta l}
        z^{\alpha-1} Z^{\ast\beta-1} \exp(il\vartheta) \right) \;,
\end{aligned}\end{equation}
if the new coordinates are denoted by $Z$ and $Z^\ast$.
From these equations, we have
\begin{equation}\begin{split}
  z 
    &= Z - \sum \beta{\cal F}_{\alpha\beta l}z^{\alpha-1} Z
         Z^{\ast\beta-1} \exp(il\vartheta) + \dots \;,
\end{split}\end{equation}
so that
\begin{equation}
  zz^\ast = ZZ^\ast + \sum(\alpha-\beta){\cal F}_{\alpha\beta l}
     Z^\alpha Z^{\ast\beta} \exp(il\vartheta) + \dots \;,
\end{equation}
where the dots indicate terms of order higher than $k$.

For the new Hamiltonian we find
\begin{equation}\label{ordkHamFkt}\begin{split}
  \phi' &= \phi-\pabl{\cF_2}{t} \\
   &= \begin{aligned}[t]
         -i\omega zz^\ast &+ \text{terms of order $<k$} \\
        &+ \sum_{\stackrel{\alpha+\beta=k}{l}} 
           \left\{\phi_{\alpha\beta l}z^\alpha z^{\ast\beta}
             + il{\cal F}_{\alpha\beta l}z^\alpha Z^\beta\right\}
           \exp(il\vartheta)\\
        &+ \text{higher-order terms}
      \end{aligned} \\
   &= \begin{aligned}[t]
         -i\omega ZZ^\ast &-i\omega\sum(\alpha-\beta)
            {\cal F}_{\alpha\beta l} Z^\alpha Z^{\ast\beta}
            \exp(il\vartheta) \\
        &+ \text{terms of order $<k$}\\
        &+ \sum
           \left\{\phi_{\alpha\beta l}+il{\cal F}_{\alpha\beta l}\right\}
              Z^\alpha Z^{\ast\beta} \exp(il\vartheta)\\         
        &+ \text{higher-order terms}
      \end{aligned} \\
   &= \begin{aligned}[t]
          -i\omega ZZ^\ast &+ \text{terms of order $<k$}\\
        &- i \sum \left\{ \left(
           \omega(\alpha-\beta)-l \right) {\cal F}_{\alpha\beta l}
           +i \phi_{\alpha\beta l} \right\} Z^\alpha Z^{\ast\beta}
           \exp (il\vartheta) \\
        &+ \text{higher-order terms} \;.
      \end{aligned}
\end{split}\end{equation}
Thus, terms of order less than $k$ remain unchanged during the
transformation, whereas if we choose
\begin{equation}
  {\cal F}_{\alpha\beta l} = -\frac {i \phi_{\alpha\beta l}}
     {\omega(\alpha-\beta)-l} \;,
  \label{ordkKoeff}
\end{equation}
the term $(\alpha\beta l)$ in (\ref{ordkErz}) cancels the term
$(\alpha\beta l)$ in the expansion of the Hamiltonian. Therefore, we
can successively eliminate terms of ever higher order without
destroying the simplifications once achieved in later steps.

Of course, the generating function (\ref{ordkErz}) must not contain
terms that make the denominator in (\ref{ordkKoeff}) vanish. Thus, we
cannot eliminate \emph{resonant terms} satisfying
$\omega(\alpha-\beta)-l=0$. If $\omega$ is irrational, only terms
having $\alpha=\beta$ and $l=0$ are resonant, so that we can transform
the Hamiltonian (\ref{NFAnsatz}) to the form
\begin{equation}
  \label{irratNF}
  \phi = -i \omega z z^\ast + \phi_2(z z^\ast)^2 + \dots + 
    \phi_k (z z^\ast)^{[k/2]} + {\cal O}((z+z^\ast)^{k+1})
\end{equation}
with arbitrarily large $k$.

If $\omega$ is rational, however, further resonant terms occur, so
that the normal form will become more complicated then
(\ref{irratNF}). These additional terms must also be kept if we want
to study the behavior of the system close to a resonance. So, let
$\omega\!\approx\!\frac{n}{m}$ with coprime integers $n$ and $m$, so
that the resonance condition $\omega(\alpha-\beta)-l=0$ reads
\begin{equation}\label{ResBed}
  n(\alpha-\beta)=m l
\end{equation}
and time-dependent resonant terms with $l\neq 0$ occur. This
time-depen\-dence can be abandoned, if we transform to a rotating
coordinate system
\begin{equation}
  Z = z \exp(in\vartheta/m)\;, \qquad\qquad
  Z^\ast = z^\ast\exp(-in\vartheta/m) \;,
\end{equation}
This transformation, which is generated by
$$ {\cal F}_2 = -Z^\ast z \exp(in\vartheta/m) \;, $$
changes resonant terms according to
\begin{equation*}\begin{split}
     z^\alpha z^{\ast\beta} \exp (il\vartheta)
  &= Z^\alpha \exp(-in\vartheta\alpha/m)
     Z^{\ast\beta} \exp(in\vartheta\beta/m)
       \exp(il\vartheta) \\
  &= Z^\alpha Z^{\ast\beta}\exp(-i\{n(\alpha-\beta)-ml\}\vartheta/m)\\
  &= Z^\alpha Z^{\ast\beta} \;.
\end{split}\end{equation*}
Thus, all resonant terms become time-independent, whereas non-resonant
terms acquire a time-dependence with period $2\pi m$. The Hamiltonian
is transformed as
\begin{equation*}\begin{split}
     \phi'
  &= \phi - \frac{\partial {\cal F}_2}{\partial t} \\
  &= \phi + i \frac{n}{m} Z^\ast z \exp(int/m) \\
  &= \phi + i \frac{n}{m} Z Z^\ast \\
  &= \begin{aligned}[t]
          -i\left(\omega -\frac{n}{m}\right) Z Z^\ast
       &+ \phi_2(Z Z^\ast)^2+\dots\\
       &+ \text{further resonant terms} \\
       &+ \text{non-resonant terms of higher order} \;,
     \end{aligned}
\end{split}\end{equation*}
that is, in the harmonic part of the Hamiltonian the frequency
$\omega$ is replaced by a small parameter
$2\varepsilon=\omega-\frac{n}{m}$ measuring the distance from the
resonance.

As we are looking for a local description of the system in a
neighborhood of the equilibrium position $z=z^\ast=0$, we can abort
the normal form transformation at a suitable $k$ and neglect
higher-order terms. This way, we get an ``idealized'' Hamiltonian that
quantitatively approximates the actual Hamiltonian close to the
equilibrium.

At the end, we return to the original coordinates $p$, $q$ or to
\emph{action-angle-coordinates} $(I,\varphi)$ given by
\begin{alignat}{2}\label{polKoord}
    p  &= \sqrt{2I}\cos\varphi \;, &q &= \sqrt{2I}\sin\varphi \;, \notag\\
    z &= \sqrt{2I}\exp(i\varphi) \;, 
      &z^\ast &=\sqrt{2I}\exp(-i\varphi) \;, \\ 
    I &= \frac{1}{2}(p^2+q^2) =  \frac{1}{2}zz^\ast \;. && \notag
\end{alignat}
We have thus obtained a selection of the most important low-order terms that
determine the behavior of the system close to the central periodic
orbit.

According to the resonance condition (\ref{ResBed}) and as $n$ and $m$
are coprime, for all resonant terms $\alpha-\beta=rm, r\in\Z$, is a
multiple of $m$, so that a resonant term has the form
\begin{equation*}\begin{split}
    z^\alpha z^{\ast\beta} 
  &= \left(\sqrt{2I}\right)^{\alpha+\beta}
     \exp\left\{i(\alpha-\beta)\varphi\right\}\\
  &= \left(\sqrt{2I}\right)^k \exp\left\{irm\varphi\right\}
\end{split}\end{equation*}
and is periodic in $\varphi$ with a period of $\frac{2\pi}{m}$. Thus,
although we started from a generic Hamiltonian, the normal form
exhibits $m$-fold rotational symmetry in a suitably chosen coordinate
system. Furthermore, all resonant terms satisfy
\begin{gather*}
  |rm| = |\alpha-\beta| \le \alpha+\beta = k \;,\\
  \alpha = \frac{1}{2}(k+rm) \in \Z \;,\\
  \beta = \frac{1}{2}(k-rm) \in \Z \;.
\end{gather*}
Thus, the normal form reads
\begin{equation}\label{NF}
  \Phi = \sum_k c_k I^k +
    \sum_k \sum_{\stackrel{0<rm\le k}{k\pm rm\text{ even}}}
       \sqrt{I}^{\,k} 
       \Big[ d_k\cos(rm\varphi) + d_k'\sin(rm\varphi) \Big] \;.
\end{equation}
As this Hamiltonian is time-independent, it is an (approximate)
constant of motion, so that all points an orbit with given initial
conditions can reach lie on a level line of the Hamiltonian. Thus, a
contour plot of the Hamiltonian will exhibit lines one will also find
in a Poincar\'e surface of section of the original Hamiltonian system.

As an example and to describe the bifurcation scenario presented in
Section \ref{BifSec}, we will now discuss the case of a fourth-order
resonance $m=4$. Up to the sixth order, the following terms turn out
to be resonant:
\begin{alignat}{2}
  \phi=&-i\left(\omega-\frac{n}{4}\right)zz^\ast 
                                         & \text{order } &2 \notag\\
    &+ \phi_2(zz^\ast)^2 + \phi_{4,0,n}z^4 + \phi_{0,4,-n}z^{\ast 4} 
                                         & &4\\
    &+ \phi_3(zz^\ast)^3 + \phi_{5,1,n}z^5z^\ast +
            \phi_{1,5,-n}zz^{\ast5} \;, \qquad
       &&6 \notag
\end{alignat}
so that the real normal form reads
\begin{equation}\label{NF4}\begin{split}
  \Phi = \quad & \varepsilon I \\
      +& a I^2+bI^2\cos(4\varphi)+b'I^2\sin(4\varphi) \\
      +& c I^3+dI^3\cos(4\varphi)+eI^3\sin(4\varphi)
\end{split}\end{equation}
with $\varepsilon = \frac{1}{2}(\omega-\frac{n}{4})$ and suitably
chosen real coefficients $a,b,b',c,d,e$. The physical meaning of these
terms will be discussed in the sequel.

\subsection{Generic Bifurcations}
\label{GenBifSec}
To lowest order, the normal form (\ref{NF4}) reads
\begin{equation}
  \Phi = \varepsilon I = \frac{\varepsilon}{2} (p^2+q^2) \;.
\end{equation}
This is a harmonic-oscillator Hamiltonian. It describes orbits which
start close to the central periodic orbit and wind around it with
frequency $\omega+\varepsilon$, or frequency $\varepsilon$ in the
rotating coordinate system.

In second order in $I$, angle-dependent terms in the normal form
occur. For arbitrary resonances, the lowest order of the normal form
containing this kind of nontrivial terms describes the generic
codimension-one bifurcations of periodic orbits as classified by Meyer
\cite{Mey70}, that is, those kinds of bifurcations that can typically
be observed if a single control parameter is varied in a system with
two degrees of freedom and without special symmetries. As was shown by
Meyer, for any order $m$ of resonance there is only one possible kind
of bifurcation, except for the case $m=4$, where there are two. In the
sequel, we are going to discuss these possibilities for $m=4$.

Up to second order in $I$, the normal form (\ref{NF4}) reads
\begin{equation}
  \Phi = \varepsilon I
      + a I^2+bI^2\cos(4\varphi)+b'I^2\sin(4\varphi) \;.
\end{equation}
Shifting the angle $\varphi$ according to
$\varphi\mapsto\varphi+\varphi_0$, we can eliminate the term
proportional to $\sin(4\varphi)$, that is, we can assume $b'=0$, so
that the normal form simplifies to
\begin{equation}\label{NFOrd2}
  \Phi = \varepsilon I + a I^2+bI^2\cos(4\varphi) \;.
\end{equation}

To find periodic orbits of the system, we have to determine the
stationary points of the normal form. The central periodic orbit we
expanded the Hamiltonian around is located at $I=0$ and does not show
up as a stationary point, because the action-angle-coordinate chart
(\ref{polKoord}) is singular there.

For $I\neq 0$, we have
\begin{equation}\begin{aligned}\label{NF2SP}
  0 &\stackrel{\textstyle !}{=} \frac{\partial \Phi}{\partial \varphi}
      = -4bI^2\sin(4\varphi) \;,\\
  0 &\stackrel{\textstyle !}{=} \frac{\partial \Phi}{\partial I}
      = \varepsilon + 2aI+2bI\cos(4\varphi) \;.
\end{aligned}\end{equation}
From the first of these equations, we get $\sin(4\varphi)=0$, that is,
$\cos(4\varphi)=\sigma\equiv\pm 1$. The second equation then yields
\begin{equation}\label{NF2I}
  I_\sigma=-\frac{\varepsilon}{2(a+\sigma b)} \;.
\end{equation}
For any choice of $\sigma$, there are four different angles $\varphi,
0\le\varphi<2\pi$, satisfying $\sin(4\varphi)=0$ and
$\cos(4\varphi)=\sigma$, corresponding to four different stationary
points in a Poincar\'e surface of section. All these stationary points
belong to the same periodic orbit, which is four times as long as the
central orbit.

\begin{figure}
  \vspace{18.0cm}
  \includegraphics{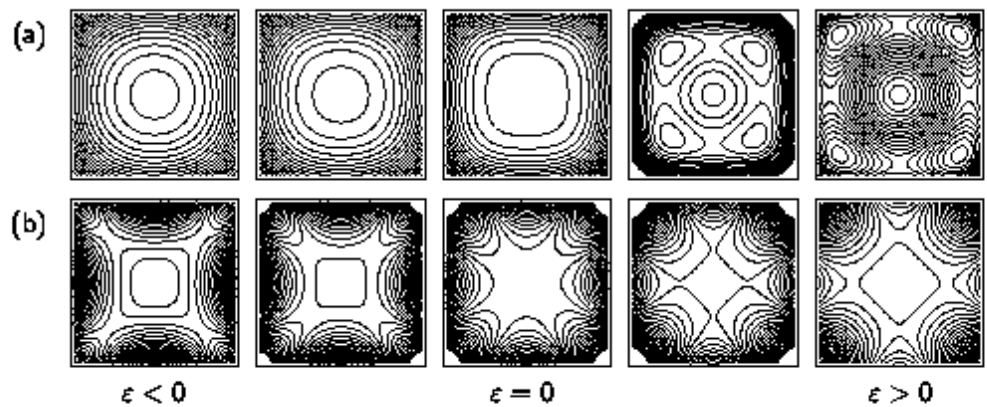}
  \caption{\label{ContPlots}
    Contour plots of the normal form (\ref{NFOrd2}) illustrating the two
    generic types of period-quadrupling bifurcations. (a)
    Island-Chain-Bifurcation, (b) Touch-and-Go-Bifurcation.}
\end{figure}

For a real periodic orbit, $I=\frac{1}{2}(p^2+q^2)$ is real and
positive. Thus, if we get a negative value for $I$ from (\ref{NF2I}),
this indicates a complex periodic orbit. The action of this orbit,
which we identify with the stationary value $\Phi(I_\sigma)$ of the
normal form, is real if $I_\sigma$ is real. Therefore, if $I_\sigma$
is real and negative, we are dealing with a ghost orbit symmetric with
respect to complex conjugation. Keeping these interpretations in mind,
we find the two possible forms of period-quadrupling bifurcations:

\begin{itemize}
\item $|a|>|b|$: Island-Chain-Bifurcation\\
  In this case, the signs of $a+b$ and $a-b$ are both equal to the sign of
  $a$. If $\sign\varepsilon=-\sign a$, both solutions $I_\sigma$ from
  equation (\ref{NF2I}) are positive, if $\sign\varepsilon=\sign a$, they
  are negative. Thus, on one hand side of the resonance, there are a stable
  and an unstable real satellite orbit. As $\varepsilon\to 0$, these orbits
  collapse onto the central periodic orbit and reappear as two complex
  satellite orbits on the other side of the resonance.
  
  Part (a) of figure \ref{ContPlots} shows a sequence of contour plots of
  the normal form, which we interprete as a sequence of Poincar\'e surface
  of section plots. If $\varepsilon<0$, we recognize a single elliptic
  fixed point at the centre of the plots, which corresponds to the stable
  central orbit. If $\varepsilon>0$, four elliptic and four hyperbolic
  fixed points appear in addition. They indicate the presence of the real
  satellite orbits. Due to these plots, the bifurcation encountered here is
  called an \emph{island-chain-bifurcation}. It was this kind of
  bifurcation which we observed in the example of section \ref{BifSec} at
  energy $E_c$.

\item $|a|<|b|$: Touch-and-Go-Bifurcation\\
  In this case, the signs of $a+b$ and $a-b$ are different, that is, at any
  given $\varepsilon$, there are a real and a complex satellite orbit. As
  $\varepsilon$ crosses 0, the real satellite becomes complex and vice
  versa.

  A sequence of contour plots for this case is shown in part (b) of
  figure \ref{ContPlots}. At any $\varepsilon$, the central elliptic
  fixed point is surrounded by four hyperbolic fixed points indicating
  the presence of an unstable real satellite. At $\varepsilon>0$, the
  fixed points are located at different angles than at
  $\varepsilon<0$, that is, it is the orbit with different $\sigma$
  which has become real. This kind of bifurcation is known as a
  \emph{touch-and-go-bifurcation}.
\end{itemize}

\subsection{Sequences of Bifurcations}
\renewcommand{\ord}[1]{{\cal O}\left({\hat{I}}^#1\right)}

In the discussion of a specific Hamiltonian system it can often be
observed that the generic bifurcations as described by Meyer occur in
organized sequences. Examples of such sequences have been discussed by
Mao and Delos \cite{Mao92} for the Diamagnetic Kepler Problem. In the
example presented in section \ref{BifSec}, we also encountered a
sequence of two bifurcations. As Sadovski\'\i, Shaw and Delos were
able to show \cite{Sad95,Sad96}, sequences of bifurcations can be
described analytically if higher order terms of the normal form
expansion are taken into account. In the sequel, we are going to use
all terms in the expansion (\ref{NF4}) up to third order in $I$. As we
did above, we can eliminate the $b'$-term if we shift $\varphi$ by a
suitably chosen constant, so that the normal form reads:
\begin{equation}\label{NFOrd3}
  \Phi=\varepsilon I+aI^2+bI^2\cos(4\varphi)+cI^3+dI^3\cos(4\varphi)
    +eI^3\sin(4\varphi) \;.
\end{equation}
It can be further simplified by canonical transformations, whereby the
transformations need to be performed up to third order in $I$ only, as
higher terms have been neglected anyway.

As a first step, we apply a canonical transformation to new
coordinates $\hat{I}$ and $\hat{\varphi}$ which is
generated by the function
\begin{equation}
  {\cal F}_2 = -\hat{I}\varphi +\frac{e}{8b}{\hat{I}}^2 \;,
\end{equation}
that is
\begin{equation}\begin{split}
  I =-\pabl{{\cal F}_2}{\varphi} &= \hat{I} \;,\\
  \hat{\varphi} = -\pabl{{\cal F}_2}{\hat{I}} 
   &= \varphi-\frac{e}{4b}\hat{I} \;.
\end{split}\end{equation}
Inserting these transformations into the normal form, we obtain up to
terms of order $\hat I^4$:
\begin{equation}
  \Phi = \varepsilon \hat{I}+a{\hat{I}}^2
        +b{\hat{I}}^2\cos(4\hat{\varphi})
        +c{\hat{I}}^3+d{\hat{I}}^3\cos(4\hat{\varphi}) \;.
\end{equation}
This expression is further simplified by another canonical
transformation generated by
\begin{equation}
  {\cal F}_2 = -\tilde{I}\hat{\varphi}
     -{\tilde{I}}^2 f(\hat{\varphi})
     - {\tilde{I}}^3 g(\hat{\varphi}) \;,
\end{equation}
where
\begin{equation}\label{fgDef}\begin{aligned}
  f(\hat{\varphi}) &= \lambda \sin(4\hat{\varphi}) \\
  g(\hat{\varphi}) &= 4\lambda^2 \left(
    \sin(4\hat{\varphi})\cos(4\hat{\varphi})-4\hat{\varphi} \right)
\end{aligned}\end{equation}
and $\lambda$ is a free parameter. Explicitly, this transformation
reads
\begin{equation}\begin{aligned}
  \hat{I}=-\pabl{{\cal F}_2}{\hat{\varphi}}
   &= \tilde{I}+{\tilde{I}}^2 f'(\hat{\varphi}) 
     + {\tilde{I}}^3 g'(\hat{\varphi}) \;,\\
  \tilde{\varphi}=-\pabl{{\cal F}_2}{\tilde{I}}
   &=\hat{\varphi} + 2\tilde{I}f(\varphi)
                   +3{\tilde{I}}^2g(\hat{\varphi}) \;,
\end{aligned}\end{equation}
from which we obtain the transformed Hamiltonian
\begin{equation}\label{NF3lambda}
  \Phi = 
     \begin{aligned}[t]
        \varepsilon \tilde{I} &+ a{\tilde{I}}^2
          +(b+4\lambda\varepsilon){\tilde{I}}^2\cos(4\tilde{\varphi})\\
       &+(c+8b\lambda){\tilde{I}}^3
          +(d+8a\lambda){\tilde{I}}^3\cos(4\tilde{\varphi})\\
       &+\ord{4}
     \end{aligned} \;.
\end{equation}
If we choose $\lambda=-\frac{d}{8a}$, we can eliminate the term
proportional to $\tilde{I}^3\cos(4\tilde{\varphi})$. Renaming
coefficients and coordinates, we finally obtain the third order
normal form
\begin{equation}\label{NF3}
  \Phi = \varepsilon I + aI^2+bI^2\cos(4\varphi)+cI^3 \;.
\end{equation}

The stationary points of this normal form except for the central
stationary point at $I=0$ are given by
\begin{equation}\begin{aligned}\label{NF3SP}
  0\stackrel{\textstyle !}{=}\pabl{\Phi}{\varphi}
   &= -4bI^2 \sin(4\varphi) \;,\\
  0\stackrel{\textstyle !}{=}\pabl{\Phi}{I}
   &= \varepsilon+2aI+2bI\cos(4\varphi)+3cI^2 \;.
\end{aligned}\end{equation}
From the first of these equations, it again follows that
\begin{gather*}
  \sin(4\varphi) = 0 \;,\\
  \cos(4\varphi) = \sigma = \pm 1 \;.
\end{gather*}
The second equation
\begin{equation*}
  \varepsilon+2(a+\sigma b)I+3cI^2 = 0
\end{equation*}
has got two solutions for any fixed $\sigma$:
\begin{equation}
  I_{\sigma\pm} = -\varrho_\sigma\pm\sqrt{\zeta+\varrho_\sigma^2}\;,
\end{equation}
where, we introduced the abbreviations
\begin{equation}
  \zeta = -\frac{\varepsilon}{3c}\;, \qquad
  \varrho_\sigma = \frac{a+\sigma b}{3c} \;.
\end{equation} 

\begin{figure}
  \vspace{18.0cm}
  \includegraphics{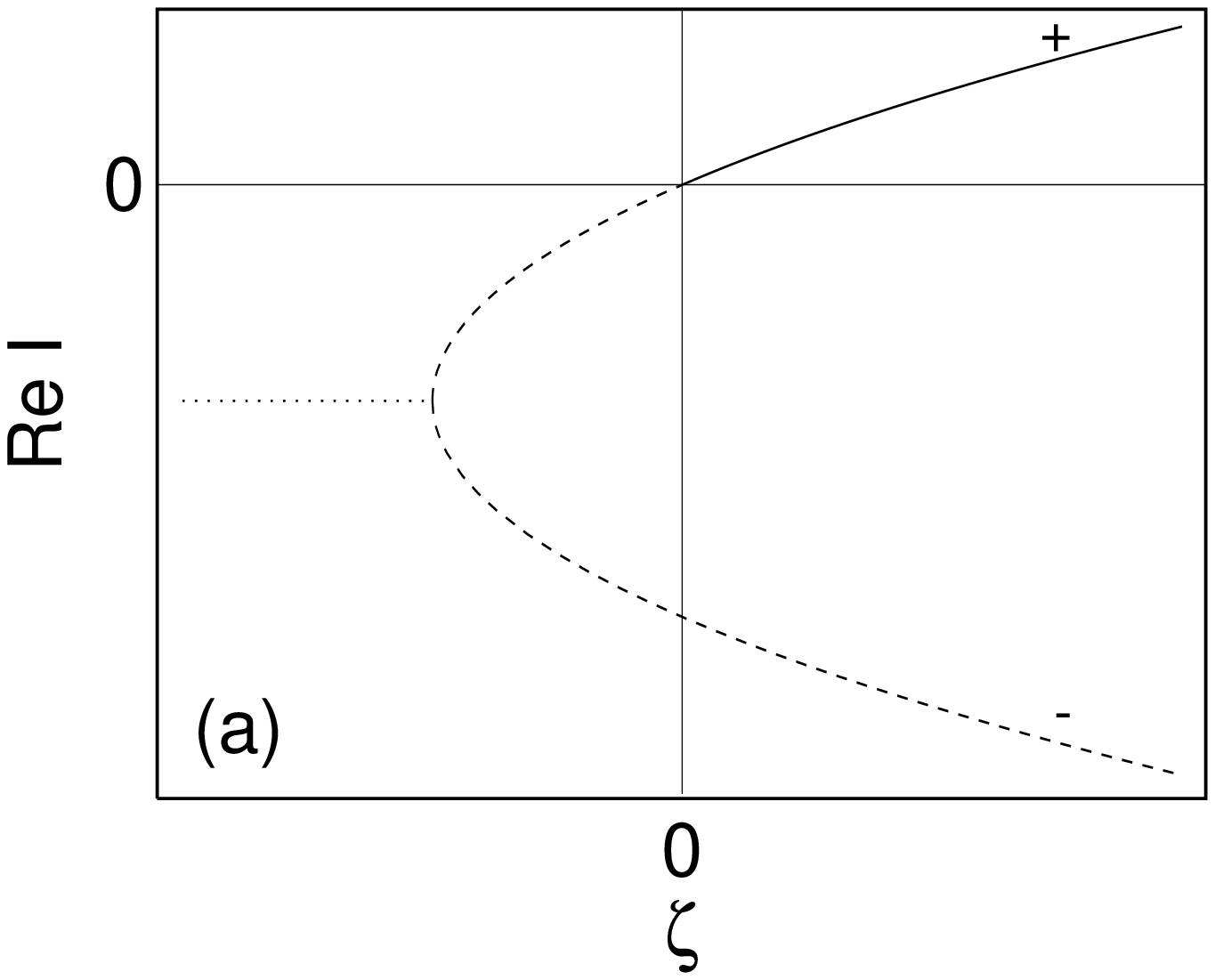}
  \includegraphics{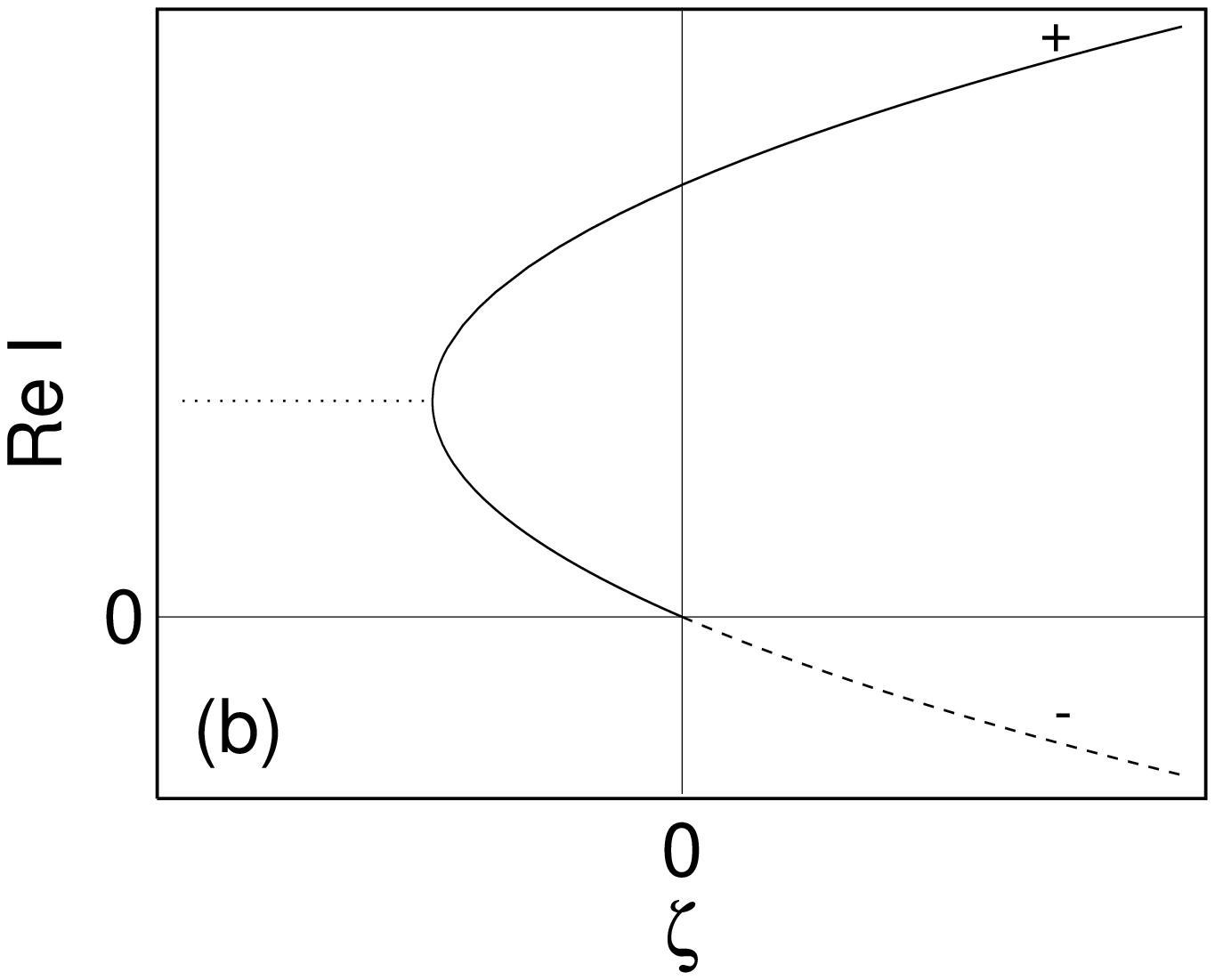}
  \caption{\label{IZeta1}
    The dependence of the radial coordinate $I_\sigma$ on $\zeta$
    illustrates the bifurcations orbits undergo. Only orbits having a
    fixed $\sigma$ are included in these plots: (a)
    $\varrho_\sigma>0$, (b) $\varrho_\sigma<0$. Solid curves: real
    orbits, dashed curves: ghost orbits symmetric with respect to
    complex conjugation, dotted curves: a pair of complex conjugate
    ghosts.}
\end{figure}

We will first discuss the behavior of the orbits with a fixed $\sigma$: The
solutions $I_{\sigma\pm}$ are real, if $\zeta>-\varrho_\sigma^2$, and they
are complex conjugates, if $\zeta<-\varrho_\sigma^2$. In figure
\ref{IZeta1}, the dependence of $I_\sigma$ on $\zeta$ is plotted for both
cases. These plots schematically exhibit the bifurcations the orbits
undergo.

\begin{itemize}
\item $\varrho_\sigma>0$\\
  In this case, $I_{\sigma -}$ is negative for $\zeta>-\varrho_\sigma^2$;
  $I_{\sigma +}$ is negative for $-\varrho_\sigma^2<\zeta<0$ and positive
  for $\zeta>0$. If we interprete this behavior in terms of periodic
  orbits, this means: The \anf{$\sigma+$}-orbit is real, if $\zeta>0$; as
  $\zeta\searrow 0$, it collapses onto the central orbit at $I=0$ and
  reappears as a ghost orbit for $\zeta<0$. At $\zeta=-\varrho_\sigma^2$,
  the \anf{$\sigma+$}-orbit collides with the \anf{$\sigma-$}-orbit, which
  has been complex up to now, and the orbits become complex conjugates of
  one another.
  
\item $\varrho_\sigma<0$\\
  In this case, $I_{\sigma +}$ is positive for $\zeta>-\varrho_\sigma^2$;
  $I_{\sigma -}$ is negative for $-\varrho_\sigma^2<\zeta<0$ and positive
  for $\zeta>0$. In terms of periodic orbits this means: The
  \anf{$\sigma-$}-orbit is complex, if $\zeta>0$, as $\zeta\searrow 0$, it
  collapses onto the central orbit and becomes real for $\zeta<0$. At
  $\zeta=-\varrho_\sigma^2$, the \anf{$\sigma-$}-orbit collides with the
  \anf{$\sigma+$}-orbit, which has been real so far, and the orbits become
  complex conjugate ghosts.
\end{itemize}

If $|a|>|b|$, the signs of $\varrho_+$ and $\varrho_-$ are equal,
whereas they are different if $|a|<|b|$. Thus, we obtain four possible
bifurcation scenarios if we take the behavior of all four satellite
orbits into account. These scenarios will be described in the sequel.

\begin{figure}
  \vspace{18.0cm}
  \includegraphics{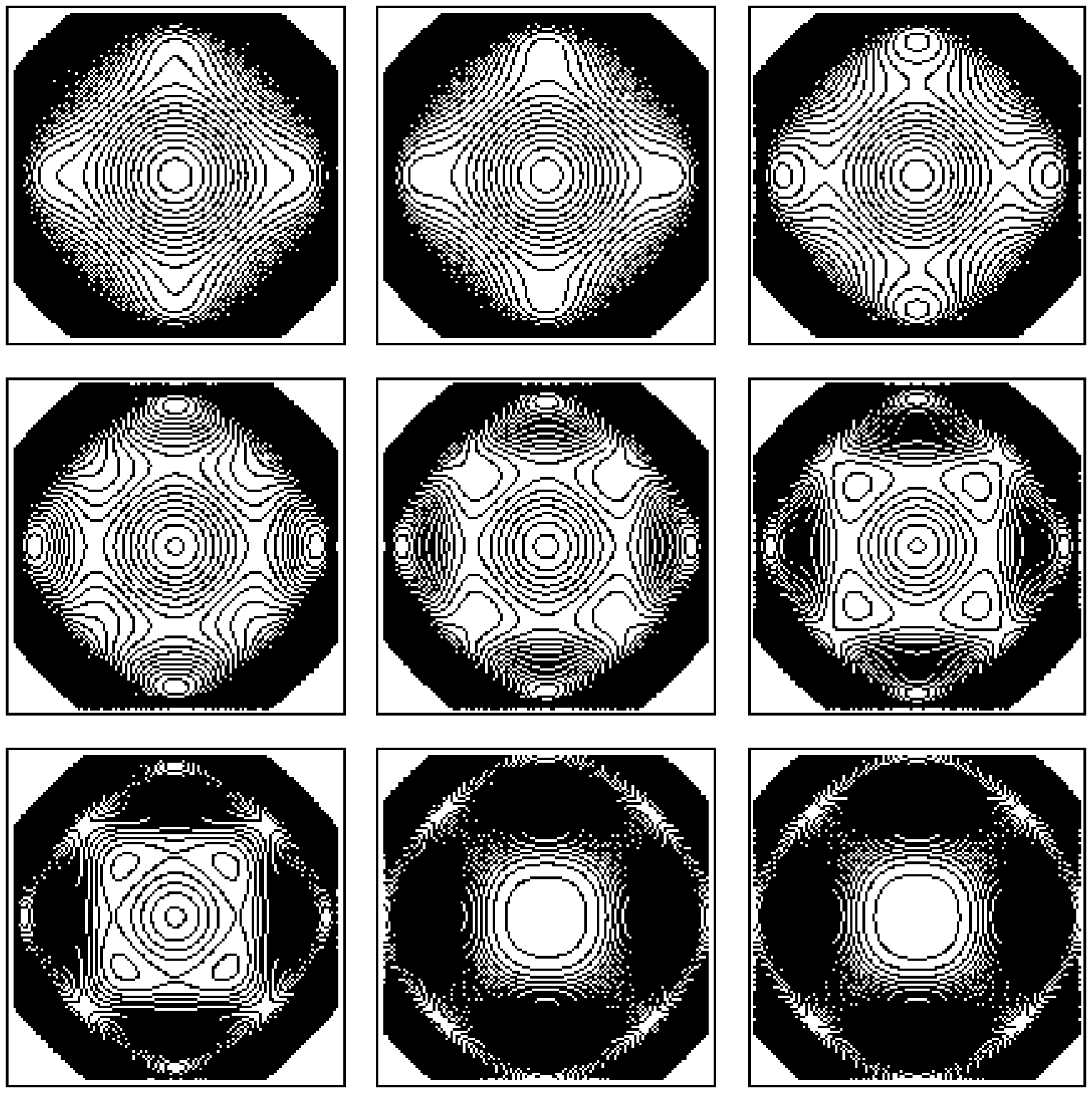}
  \caption{\label{nongen_contour}
    Contour plots of the normal form (\ref{NF3}) illustrating the
    bifurcation scenario of case 1.}
\end{figure}

\begin{figure}
  \vspace{18.0cm}
  \includegraphics{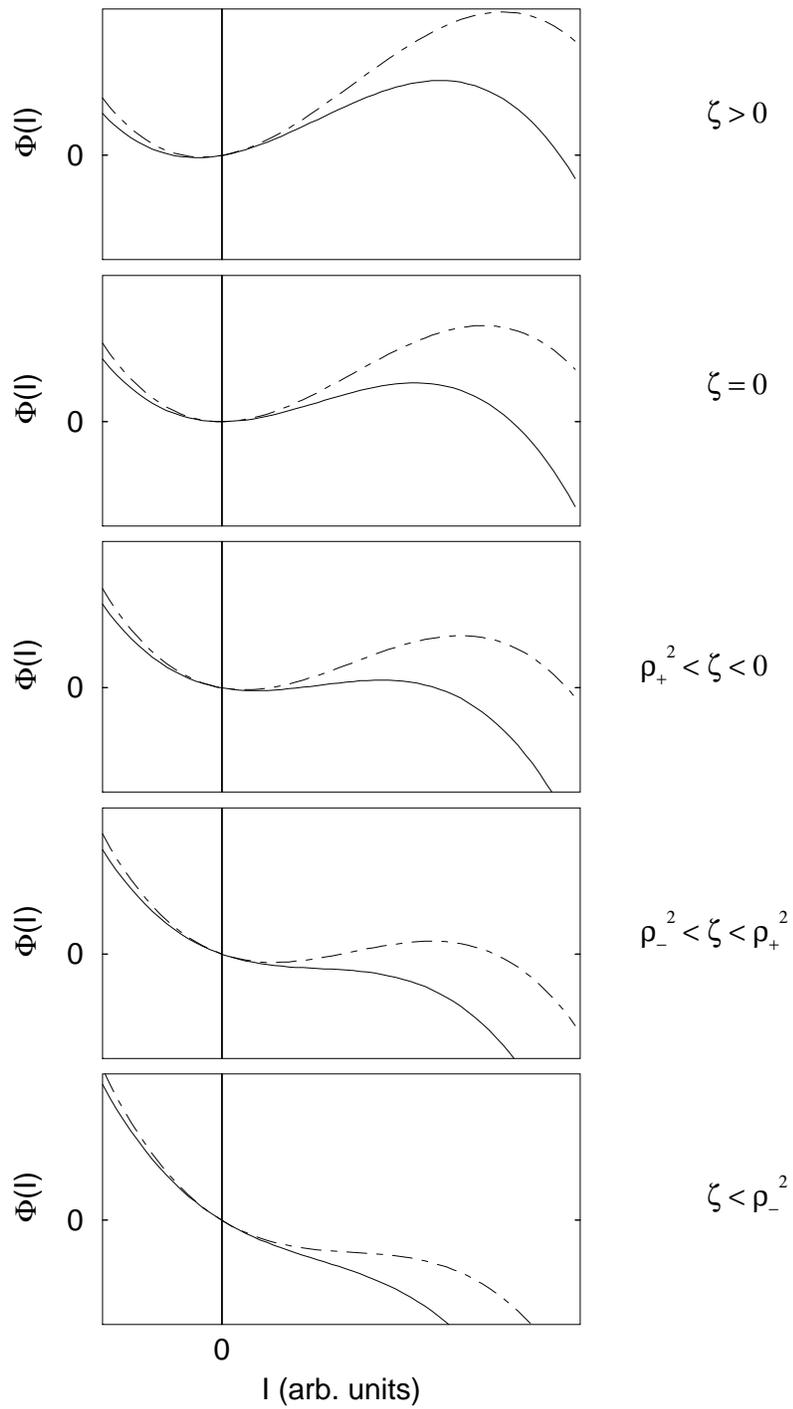}
  \caption{\label{NFI1}
    Plots of the normal form illustrating the bifurcation scenario of case
    1. Solid curve: $\sigma=+1$, dashed-dotted curve: $\sigma=-1$.}
\end{figure}

\begin{figure}
  \vspace{18.0cm}
  \includegraphics{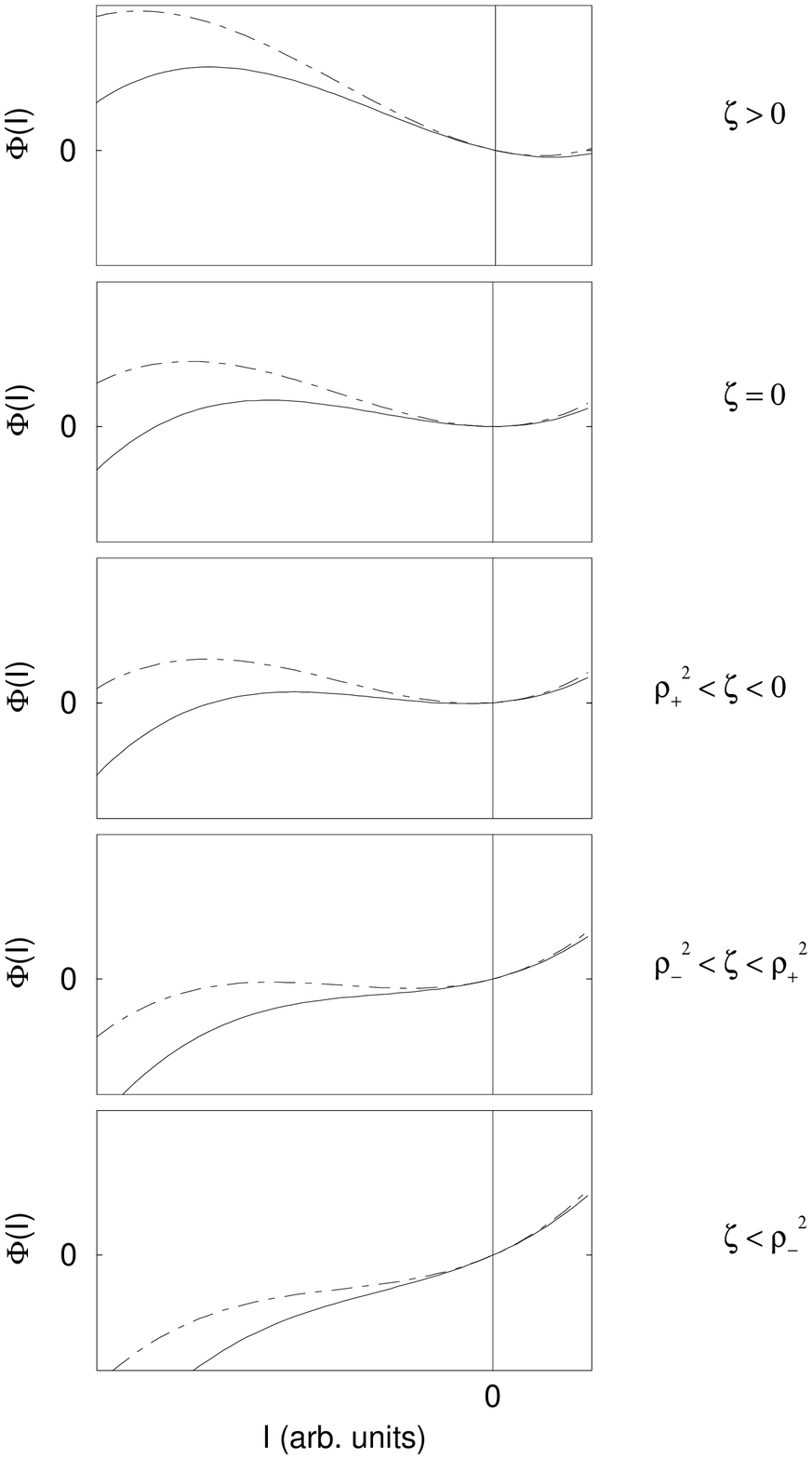}
  \caption{\label{NFI2}
    Plots of the normal form illustrating the bifurcation scenario of case
    2.}
\end{figure}

\begin{figure}
  \vspace{18.0cm}
  \includegraphics{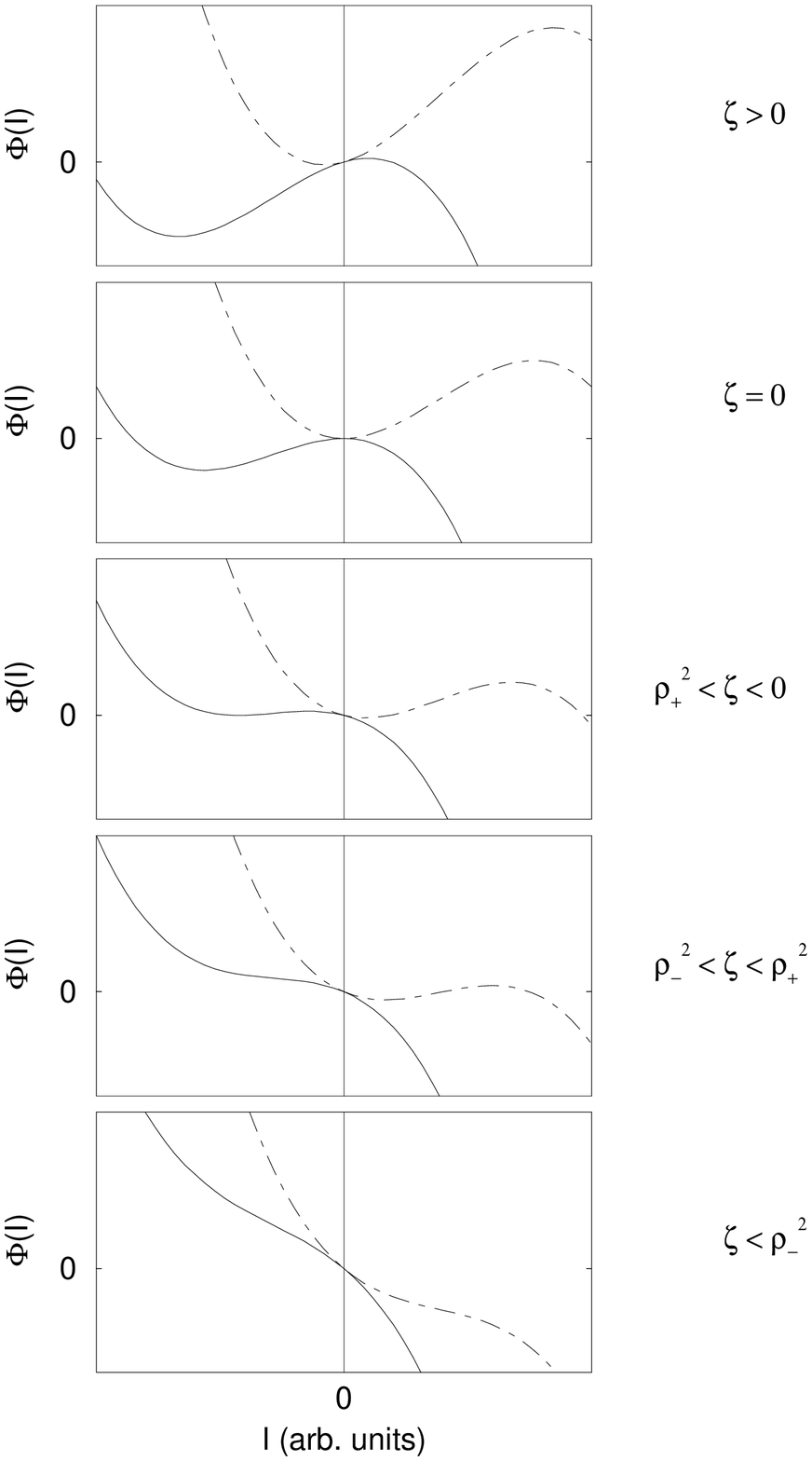}
  \caption{\label{NFI3}
    Plots of the normal form illustrating the bifurcation scenario of case
    3.}
\end{figure}

\begin{figure}
  \vspace{18.0cm}
  \includegraphics{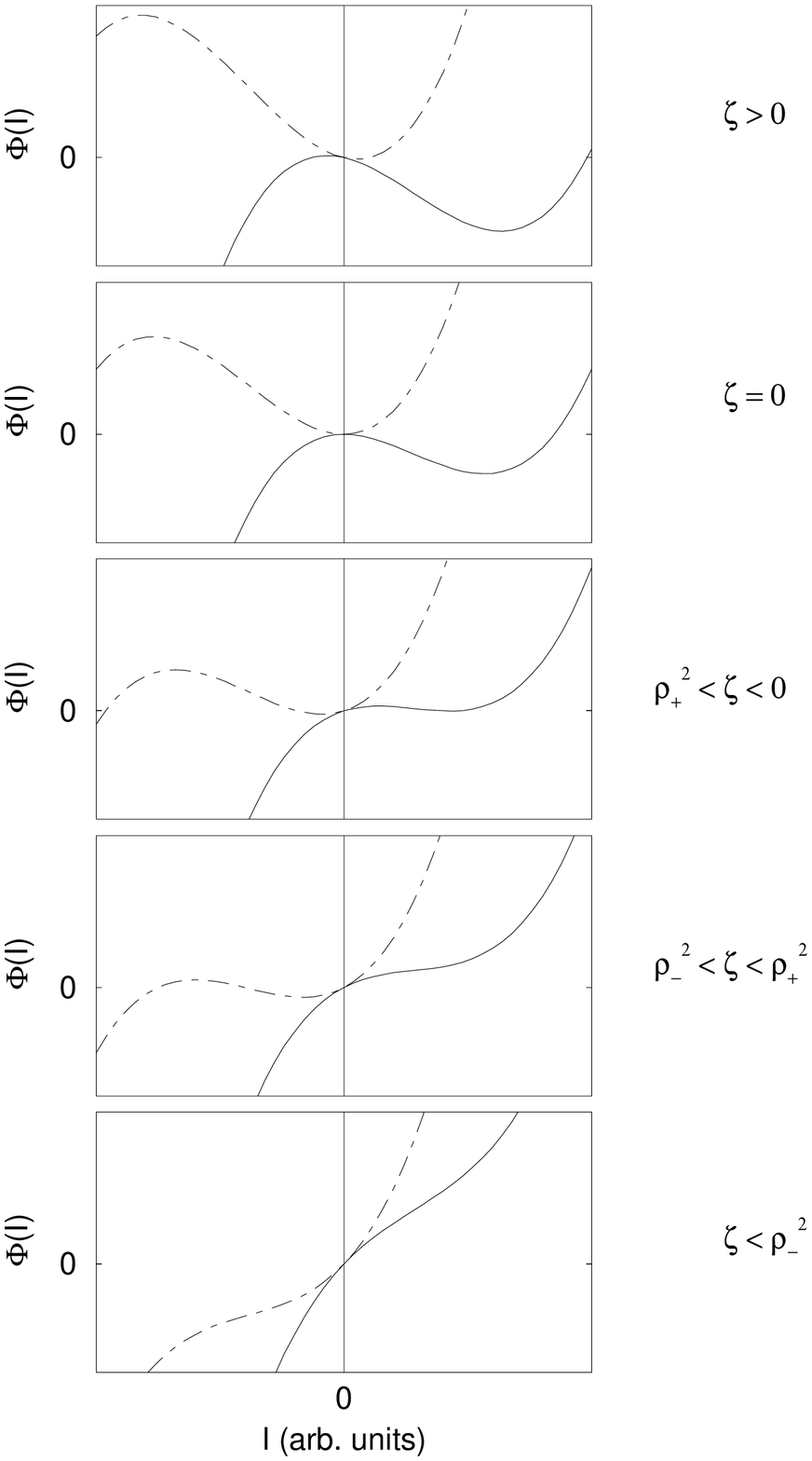}
  \caption{\label{NFI4}
    Plots of the normal form illustrating the bifurcation scenario of case
    4.}
\end{figure}

\begin{figure}
  \vspace{18.0cm}
  \includegraphics{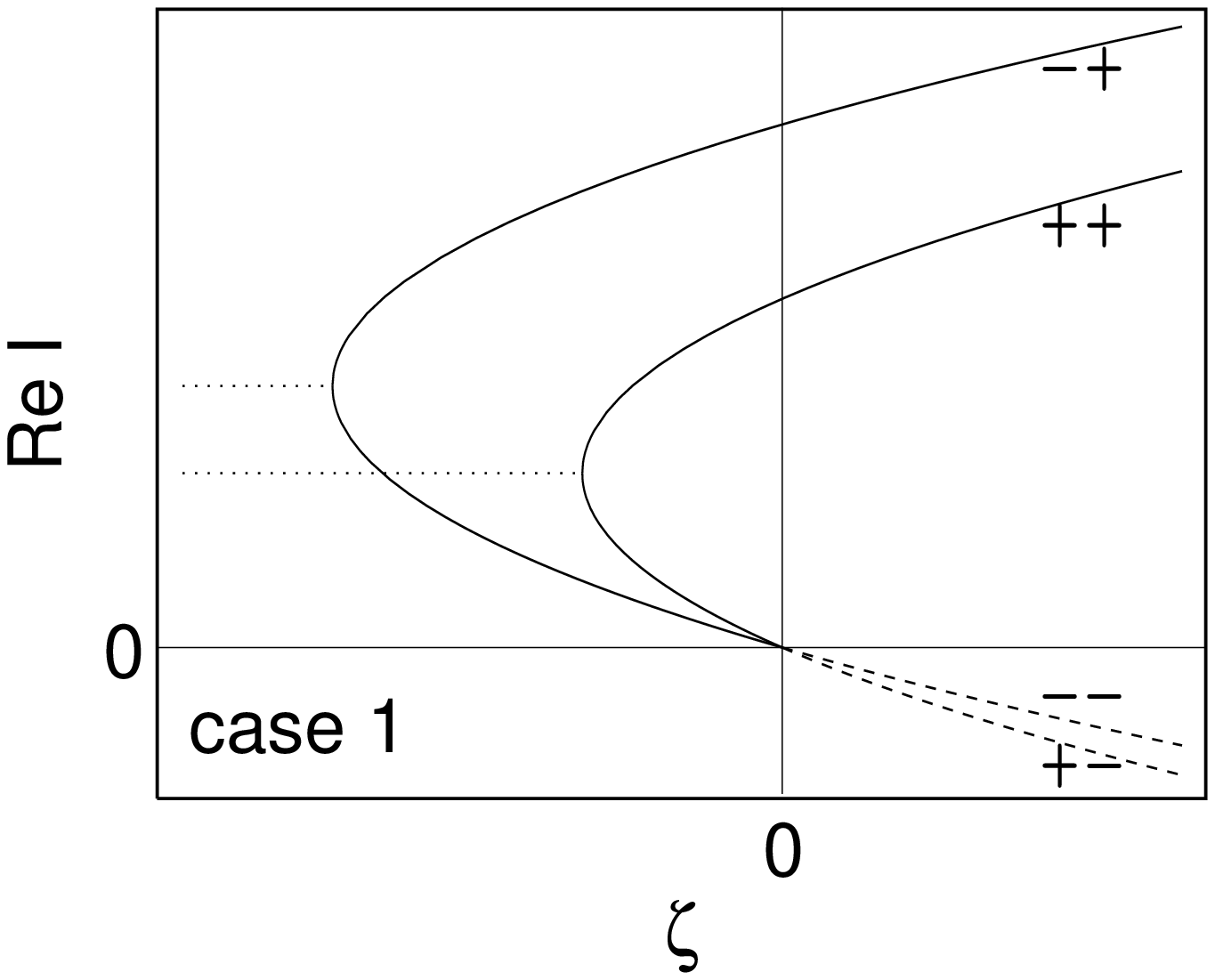}
  \includegraphics{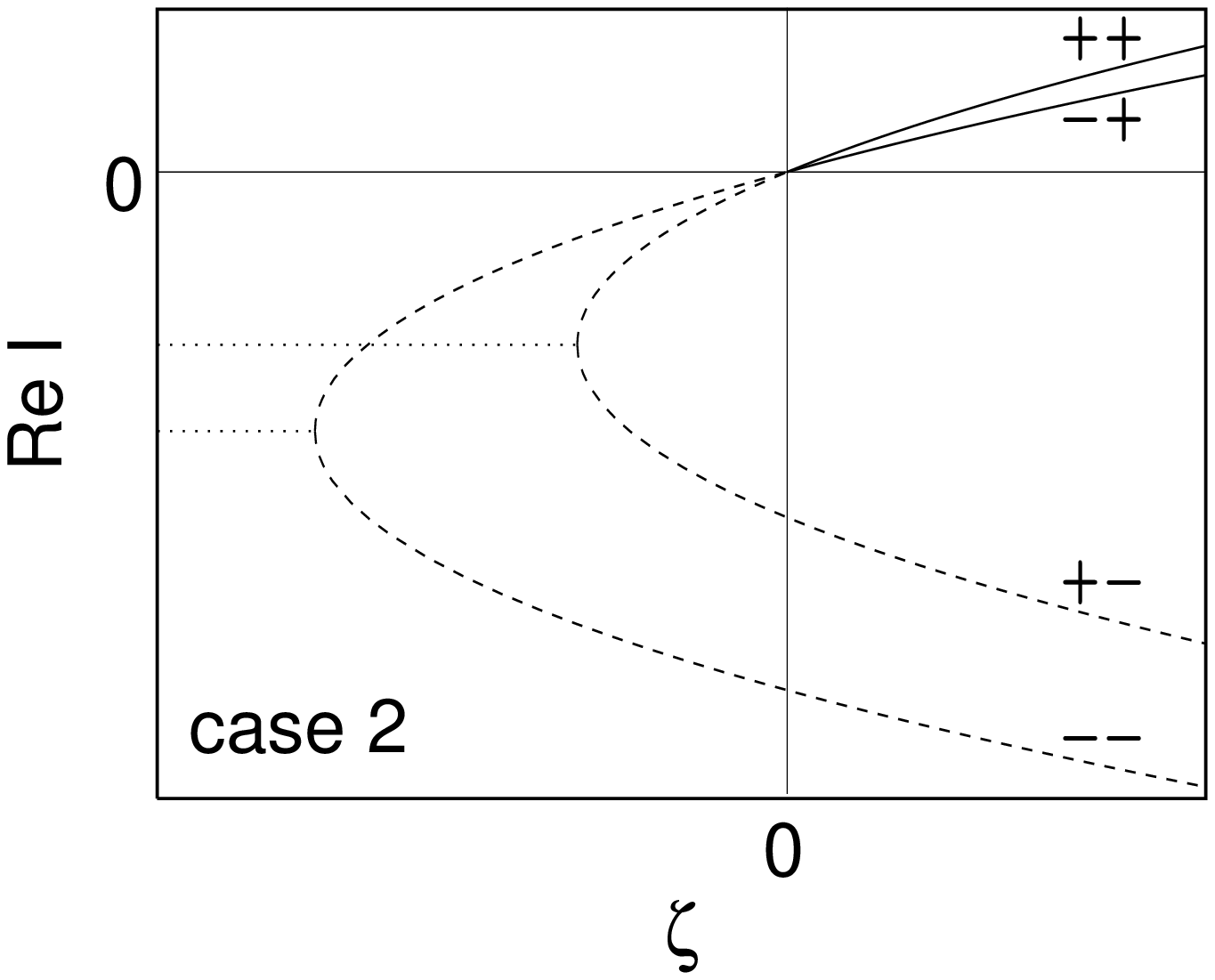}
  \includegraphics{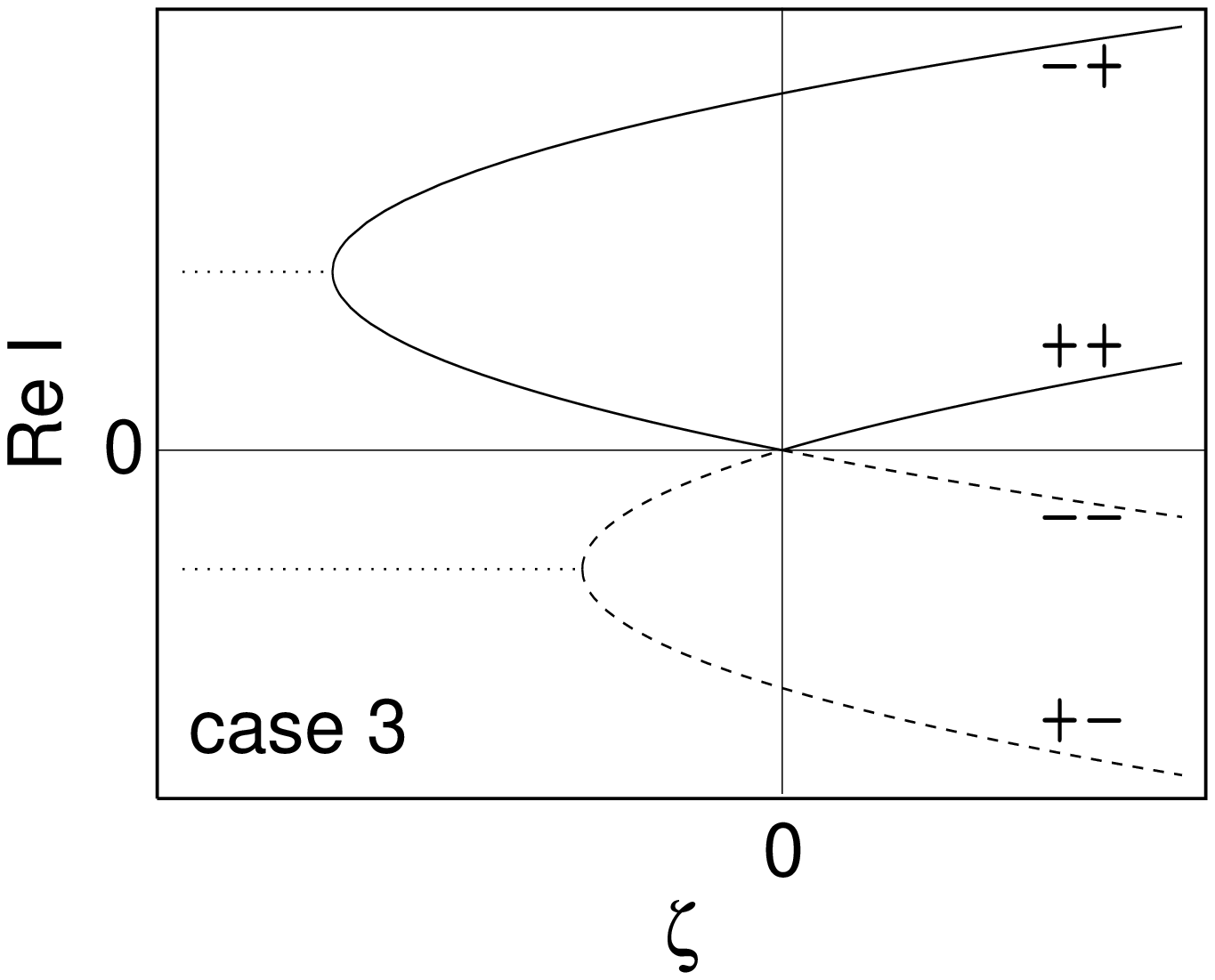}
  \includegraphics{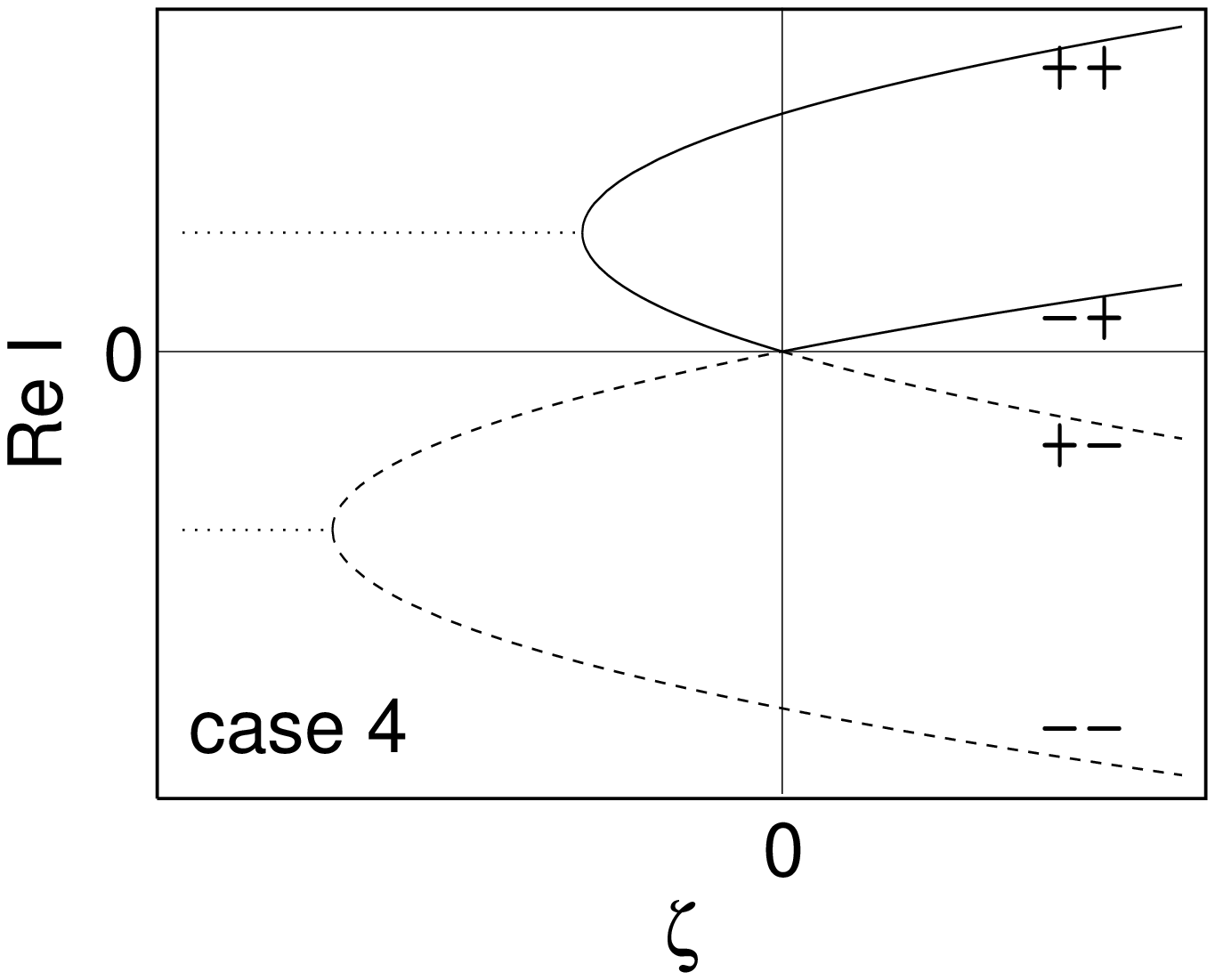}
  \caption{\label{IZeta2}
    The four complete bifurcation scenarios the normal form
    (\ref{NF3}) can describe.}
\end{figure}

\begin{enumerate}
\item $|a|>|b|,\quad\varrho_{\pm}<0$\\
  The orbits \anf{$+-$} and \anf{$--$} are ghosts if $\zeta>0$. At
  $\zeta=0$ they collide to form an island-chain-bifurcation with the
  central periodic orbit and become real if $\zeta<0$. At
  $\zeta=-\varrho_\sigma^2$, the real satellite \anf{$\sigma-$} collides
  with the real \anf{$\sigma+$}-orbit, and they become complex conjugate
  ghosts.
  
  A sequence of contour plots of the normal form describing this scenario
  is given in figure \ref{nongen_contour}. The plots are arranged in order
  of decreasing $\zeta$.  If $\zeta>0$, the central orbit is surrounded by
  a chain of four elliptic and four hyperbolic fixed points, representing a
  stable and an unstable orbit of quadruple period.  At $\zeta=0$, another
  pair of quadruple-period orbits is created. As $\zeta$ decreases further,
  two subsequent tangent bifurcations occur, each of them destroying one
  orbit from the inner and from the outer island chain. In effect, all
  satellite orbits have gone, giving the overall impression as if a single
  period-quadrupling bifurcation had destroyed the outer island chain,
  whereas in fact a complicated sequence of bifurcations has taken place.

  In the remaining cases we are going to discuss, bifurcations of ghost
  orbits occur which cannot be seen in contour plots. Therefore, we have to
  resort to a different kind of presentation. In figure \ref{NFI1} we plot
  the value of the normal form $\Phi$, depending on the action coordinate
  $I$, for both $\sigma = +1$ and $\sigma = -1$. In these plots, a periodic
  orbit corresponds to a stationary point of $\Phi(I)$. If the stationary
  point occurs at a positive value of $I$, it indicates the presence of a
  real orbit, whereas a stationary point at a negative $I$ corresponds to a
  ghost orbit symmetric with respect to complex conjugation. Asymmetric
  ghost orbits correspond to stationary points at complex $I$ and are
  therefore invisible.
  
  The bifurcation scenario described above manifests itself in the plots as
  follows: If $\zeta>0$, there are stationary points at positive and
  negative values of $I$ for both $\sigma=+1$ and $\sigma=-1$. As $\zeta$
  becomes negative, the stationary points at negative $I$ simultaneously
  cross the $\Phi$-axis and move to positive values of $I$, indicating the
  occurence of an island-chain-bifurcation and the appearance of two real
  orbits. As $\zeta$ decreases further, the two stationary points of the
  \anf{$\sigma=+1$}-curve collide and disappear as the two orbits vanish in
  a tangent bifurcation. Subsequently, the same happens to the
  \anf{$\sigma=-1$}-orbits.

\item $|a|>|b|,\quad\varrho_{\pm}>0$\\
  The orbits \anf{$++$} and \anf{$-+$} are real if $\zeta>0$. As
  $\zeta\searrow 0$, they simultaneously collapse onto the central periodic
  orbit and become ghosts, that is, at $\zeta=0$ an
  island-chain-bifurcation takes place. For any $\sigma$, the complex
  \anf{$\sigma+$}- and \anf{$\sigma-$}-orbits collide at
  $\zeta=-\varrho_\sigma^2$ and become complex conjugates.  This sequence
  of events is depicted in figure \ref{NFI2}.
  
\item $|a|<|b|,\quad\varrho_-<0<\varrho_+$\\
  If $\zeta>0$, the orbit \anf{$++$} is real, whereas \anf{$--$} is
  complex. As $\zeta\searrow 0$, these orbits collapse onto the central
  orbit and form a touch-and-go-bifurcation.  In the plots of figure
  \ref{NFI3}, this bifurcation manifests itself in two stationary points
  simultaneously crossing the $\Phi$-axis from opposite sides. At
  $\zeta=-\varrho_+^2$, the \anf{$++$}-orbit, which is complex now,
  collides with the complex \anf{$+-$}-orbit, and they become complex
  conjugate ghost orbits. Similarly, the real orbits \anf{$--$} and
  \anf{$-+$} become complex conjugates in a collision at
  $\zeta=-\varrho_-^2$.
  
\item $|a|<|b|,\quad\varrho_+<0<\varrho_-$\\
  This case is similar to the preceding. Following the
  touch-and-go-bi\-fur\-ca\-tion at $\zeta=0$, the disappearences of the
  real and the ghost orbits now occur in reversed order (see figure
  \ref{NFI4}).
\end{enumerate}
Figure \ref{IZeta2} summarizes the four bifurcation scenarios described
above. As in figure \ref{IZeta1}, we plot the values of $I$ where the
stationary points occur for different $\zeta$, so that the sequence of
bifurcations becomes visible in a single plot.

The scenario called case 2 above is already rather similar to the
situation discussed in section \ref{BifSec}. However, in our example
we observed only one of the two ghost orbit bifurcations, and there is
no actual periodic orbit corresponding to the \anf{$--$}-orbit of the
normal form. To obtain a more accurate description of the bifurcation
phenomenon, we adopt a slightly different normal form by setting
$\lambda=\frac{d-c}{8(b-c)}$ in (\ref{NF3lambda}). After renaming, we
obtain the modified normal form
\begin{equation}\label{NF3Var}
  \Phi = \varepsilon I+a I^2+b I^2\cos(4\varphi)
      + c I^3 (1+\cos(4\varphi)) \;.
\end{equation}
The stationary-point equations read
\begin{equation}\begin{aligned}
  0\stackrel{\textstyle !}{=} \pabl{\Phi}{\varphi}
    &= -4I^2(b+cI)\sin(4\varphi) \;,\\
  0\stackrel{\textstyle !}{=} \pabl{\Phi}{I}
    &= \varepsilon+2aI+2bI\cos(4\varphi)
       +3cI^2(1+\cos(4\varphi)) \;.
\end{aligned}\end{equation}
As above, it follows that
\begin{gather*}
  \sin(4\varphi)=0 \;,\\
  \cos(4\varphi) = \sigma = \pm 1 \;,
\end{gather*}
and
\begin{equation}\label{NF3aSP}
  \varepsilon+2(a+\sigma b)I+3cI^2(1+\sigma) = 0 \;.
\end{equation}

If $\sigma=+1$, this agrees with equation (\ref{NF3SP}) as obtained
from the third-order normal form discussed above and thus yields the
sequence of period quadrupling and isochronous bifurcation as found
above. If $\sigma=-1$, however, the third-order term vanishes, so that
there is only one further satellite orbit described by the normal form
which is directly involved in the period quadrupling.  To put it more
precisely, the stationary points of the normal form for $\sigma=+1$
occur at
\begin{equation*}
  I_\pm= -\frac{a+b}{6c} \pm 
    \sqrt{-\frac{\varepsilon}{6c}+\left(\frac{a+b}{6c}\right)^2}
\end{equation*}
and for $\sigma=-1$ at
\begin{equation*}
  I_{-1} = -\frac{\varepsilon}{2(a-b)} \;.
\end{equation*}

\begin{figure}
  \vspace{18.0cm}
  \includegraphics{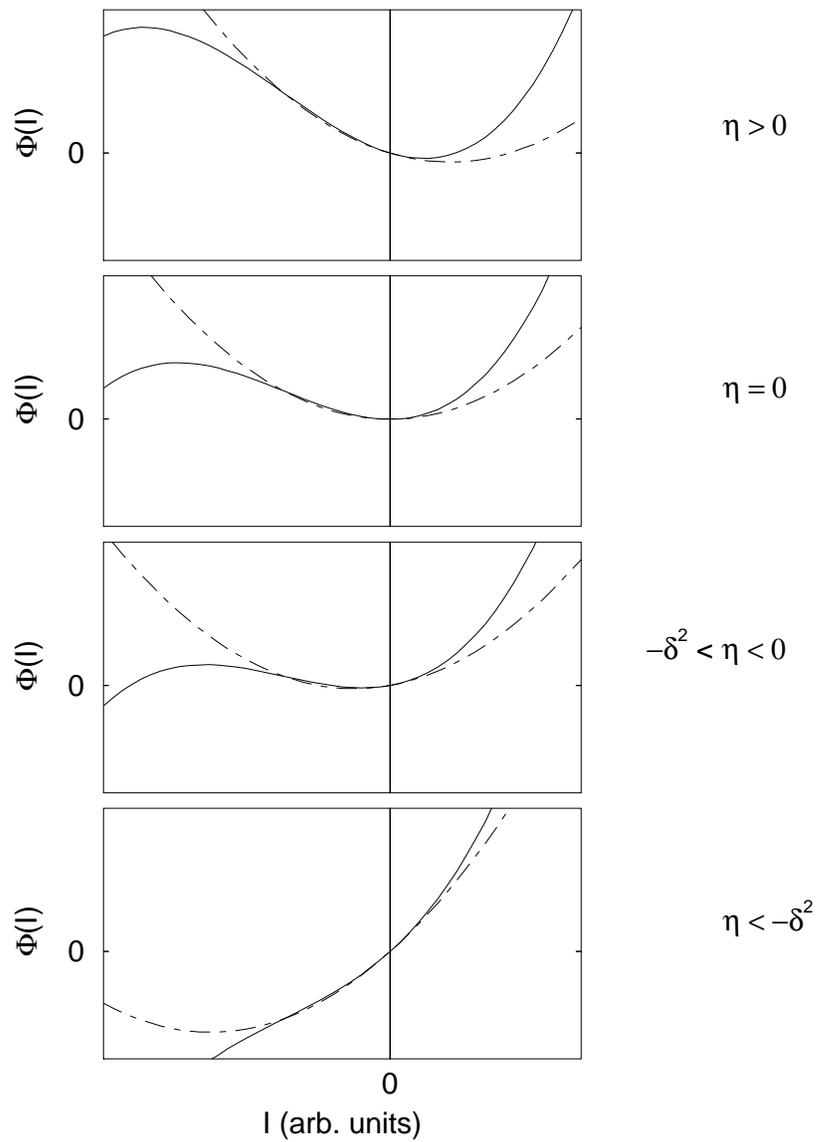}
  \caption{\label{NFIVar}
    Plots of the normal form illustrating the bifurcation scenario
    described by the normal form (\ref{NF3Var}).}
\end{figure}

\begin{figure}
  \vspace{18.0cm}
  \includegraphics{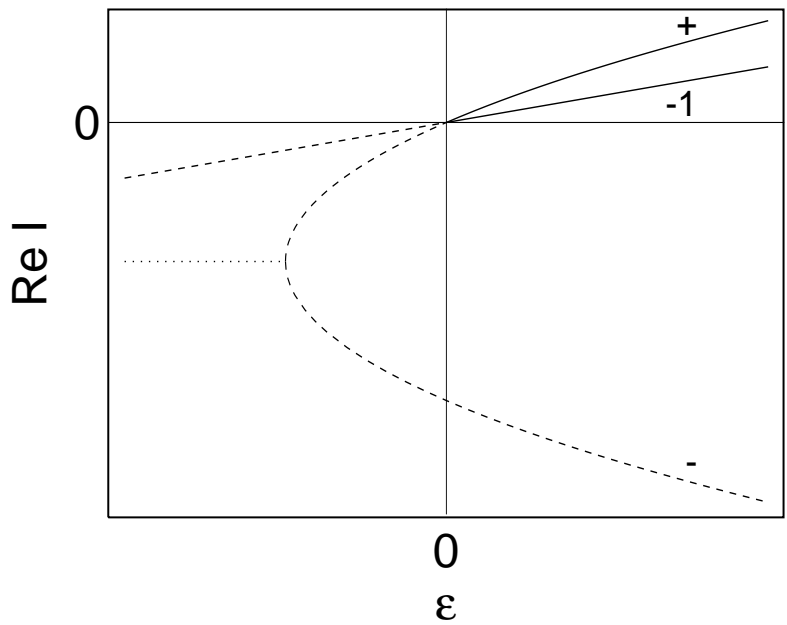}
  \caption{\label{IZetaVar}
    Bifurcation scenario described by the normal form (\ref{NF3Var})
    in the case $|a|>|b|$ and $c<0$.}
\end{figure}

From now on, we will assume $|a|>|b|$ and $c<0$. As can be shown in a
discussion similar to the above, this is the only case in which an
island-chain-bifurcation occurs at $\varepsilon=0$ with the real satellites
existing for positive $\varepsilon$ as we need to describe our example
situation from section \ref{BifSec}. The bifurcation scenario described by
the normal form (\ref{NF3Var}) in this case is shown schematically in
figures \ref{NFIVar} and \ref{IZetaVar}. The sequence of an
island-chain-bifurcation at $\varepsilon=0$ and a ghost orbit bifurcation
at some negative value of $\varepsilon$ can easily be seen to agree with
the bifurcation scenario described in section \ref{BifSec}. Furthermore,
with the help of the bifurcations the orbits undergo we can identify
individual periodic orbits with stationary points of the normal form as
follows: The central periodic orbit corresponds to the stationary point at
$I=0$ by construction. The stationary point labelled as \anf{$-1$} collides
with the origin at $\varepsilon=0$, but does not undergo any further
bifurcations. It can thus be identified with the unstable satellite orbit.
Finally, the stationary points \anf{$+$} and \anf{$-$} agree with the
stable satellite orbit and the additional ghost orbit, respectively.

Under the above assumptions, we can write
\begin{equation}
  I_\pm = -c^{-1/3}\left( \delta\pm\sqrt{\eta+\delta^2} \right)
\end{equation}
with the abbreviations
\begin{equation}\label{EtaDeltaDef}\begin{gathered}
  \eta = -\frac{\varepsilon}{6c^{1/3}} \;,\\
  \delta = \frac{a+b}{6c^{2/3}} \;.
\end{gathered}\end{equation}
From the decomposition
\begin{equation*}
  \Phi = \left( \frac{1}{3}I + \frac{a+b}{18c} \right) \pabl{\Phi}{I}
   -4c^{1/3}\left(\eta+\delta^2\right)I+2\eta\delta
  \text{\qquad if } \cos(4\varphi)=+1 \;,
\end{equation*}
which can be derived by a polynomial division, we then obtain the
actions of periodic orbits as
\begin{equation}\label{NF3Wirk}\begin{gathered}
  \begin{split}
    \Phi_\pm &= \Phi(I_\pm,\sigma=+1) \\ 
      &= -4c^{1/3}\left(\eta+\delta^2\right)I_\pm+2\eta\delta \\
      &= +4\left(\eta+\delta^2\right)
          \left(\delta\pm\sqrt{\eta+\delta^2}\right)+2\eta\delta\;,
  \end{split}\\
  \Phi_{-1} = \Phi\left(I_{-1},\sigma=-1\right) 
    = -\frac{\varepsilon^2}{4(a-b)} \;,\\
  \Phi_0 = \Phi(I=0) = 0 \;.
\end{gathered}\end{equation}

Furthermore, we shall need the Hessian Determinants of the action
function at the stationary points. We can calculate them in an
arbitrary coordinate system in principle. However, the polar
coordinate system $(I,\varphi)$ is singular at the position of the
central periodic orbit, so that we cannot calculate a Hessian
determinant there. Thus, we will use Cartesian coordinates. Using the
transformation equations (\ref{polKoord}) and the relation
\begin{equation*}
  \cos(4\varphi) = \cos^4\varphi-6\cos^2\varphi\sin^2\varphi
    +\sin^4\varphi\;,
\end{equation*}
we can express the normal form in Cartesian coordinates
\begin{multline}
  \Phi= \frac{\varepsilon}{2}(p^2+q^2)+\frac{a}{4}(p^4+2p^2q^2+q^4) \\
   +\frac{b}{4}(p^4-6p^2q^2+q^4)+\frac{c}{4}(p^6-p^4q^2-p^2q^4+q^6)\;.
\end{multline}
This yields the Hessian determinants
\begin{equation}\begin{split}
 \Hess \Phi =& \Phi_{pp}\Phi_{qq}-\Phi_{pq}\Phi_{qp} \\
  =& \left\{\varepsilon+3(a-b)p^2+(a-3b)q^2
            +\frac{c}{2}(15p^4-6p^2q^2-q^4)\right\} \times \\
   &\quad
     \left\{\varepsilon+(a-3b)p^2+3(a+b)q^2
            -\frac{c}{2}(p^4+6p^2q^2-15q^4)\right\} \\
   & -4p^2q^2\left\{a-3b-c(p^2+q^2)\right\}^2
\end{split}\end{equation}
If we pick $p=q=0$ on the central periodic orbit, $p=0,
q=\sqrt{2I_\pm}$, that is, $\varphi=0$, for $\sigma=+1$, and
$p=q=\frac{1}{\sqrt{2}}\sqrt{2I_{-1}} =\sqrt{I_{-1}}$, that is,
$\varphi=\frac{\pi}{4}$, for $\sigma=-1$, we finally obtain Hessian
determinants at the periodic orbits:
\begin{equation}\label{NF3Dets}\begin{gathered}
  \Hess_\pm = \left\{\varepsilon+2(a-3b)I_\pm-2cI_\pm^2\right\}
      \left\{\varepsilon+6(a+b)I_\pm+30cI_\pm^2\right\} \;,\\
  \Hess_{-1} = \left\{\varepsilon+4aI_{-1}+4cI_{-1}^2\right\}^2
      -4I_{-1}^2\left\{a-3b-2cI\right\}^2 \;,\\
  \Hess_0 = \varepsilon^2 \;.
\end{gathered}\end{equation}

We have now found an analytic description of the bifurcation scenario
we are discussing, and we have evaluated stationary values and Hessian
determinants which we will relate to classical parameters of the
orbits. We will now go over to the constuction of a uniform
approximation.

\section{Uniform approximation}
\label{UnifLsgSec}

\subsection{General derivation of the uniform approximation}
We need to calculate the collective contribution of all orbits
involved in the bifurcation scenario to the density of states. In the
derivation of the integral representation (\ref{IntDarst}) of the
uniform approximation, we take the method used by Sieber \cite{Sie96}
as a guideline.

We use the semiclassical Green's function
(\ref{sclGreen}) as a starting point and include the contribution of a
single orbit only:
\begin{equation}
  G(\vec{x}'\vec{x},E) = \frac{1}{i\hbar\sqrt{2\pi i\hbar}}
   \sqrt{|D|}\exp\left\{ \frac{i}{\hbar}S(\vec{x}'\vec{x},E)-
      i\frac{\pi}{2}\nu\right\} \;.
\end{equation}
Here, $S$ denotes the action of the periodic orbit, $\nu$ its
Maslov-index, and
\begin{equation*}
  D = \det
   \begin{pmatrix}
     \frac{\partial^2S}{\partial\vec{x}'\partial\vec{x}} & 
     \frac{\partial^2S}{\partial\vec{x}'\partial E} \\ [1ex]
     \frac{\partial^2S}{\partial E\partial\vec{x}} &
     \frac{\partial^2S}{\partial E^2}
   \end{pmatrix} \;.
\end{equation*}
As in section \ref{NFBifSec}, we introduce configuration space
coordinates $(y,z)$ so that $z$ is measured along the periodic orbit
and increases by $2\pi$ within each circle and $y$ measures the
distance form the orbit. We then have
\begin{equation*}\begin{split}\label{TrGForm}
  \Tr G &=\int d^2x' d^2x\, \delta(\vec{x}'-\vec{x})
           G(\vec{x}'\vec{x},E)\\ 
    &= \frac{1}{i\hbar\sqrt{2\pi i\hbar}} \int d^2x'd^2x\,
      \delta(z'-z)\delta(y'-y) \sqrt{|D|} \\
    &\hspace{4cm}\times
      \exp\left\{\frac{i}{\hbar}S(\vec{x}'\vec{x},E)-i\frac{\pi}{2}\nu
          \right\} \\
    &= \frac{1}{i\hbar m\sqrt{2\pi i\hbar}} \int dy'dz\,dy\,
      \delta(y'-y) \sqrt{|D|} \\
    &\hspace{4cm}\times
      \exp\left\{\frac{i}{\hbar}S(\vec{x}'\vec{x},E)-i\frac{\pi}{2}\nu
          \right\} \Bigg|_{z'=z+2\pi m} \;.
\end{split}\end{equation*}
In the last step, the $z'$-integration has been performed. Close to
an $\frac{n}{m}$-resonance, we regard $m$ periods of the bifurcating
orbit as the central periodic orbit, so that the $z$-integration
extends over $m$ primitive periods, although it should only extend
over one. This error is corrected by the prefactor $\frac{1}{m}$.

From its definition (\ref{SDef}), the action integral $S(\vec x'\vec x, E)$
obviously satisfies
\begin{equation}
  \pabl S{\vec x'} =  \vec p' \;, \qquad\qquad
  \pabl S{\vec x} =  -\vec p \;.
\end{equation}
We can thus regard $S$ as the coordinate representation of the generating
function of the $m$-traversal Poincar\'e map. At a resonance, however, the
Poincar\'e map is approximately equal to the identity map whose generating
function does not possess a representation depending on old and new
coordinates. Thus, we go over to a coordinate-momentum-representation. To
this end, we substitute the integral representation
\begin{equation*}
  \delta(y'-y) = \frac{1}{2\pi\hbar}\int_{-\infty}^{+\infty} dp_y'\,
    \exp\left\{\frac{i}{\hbar}p'_y(y-y')\right\}
\end{equation*}
into (\ref{TrGForm}) and evaluate the $y'$-integration using the
stationary-phase approximation. The stationarity condition reads
\begin{equation}\label{statBed}
  \pabl{S}{y'}-p_y'=0 \;,
\end{equation}
so that we obtain
\begin{multline}\label{TrGa}
  \Tr G = \frac{1}{2\pi im\hbar^2}\int dy\,dz\,dp_y'
    \sqrt{|D|\big|_{\text{sp}}} \\
    \times 
    \exp\left\{\frac{i}{\hbar}\left(\hat{S}+yp_y'\right)
       -i\frac{\pi}{2}\hat{\nu} \right\}
    \frac{1}{\sqrt{\left.\left|\frac{\partial^2S}
                  {\partial y'{}^2}\right|\right|_{\text{sp}}}} \;.
\end{multline}
Here,
\begin{equation}
  \hat{S}(z'p_y'zy,E) = S(z'y'z\,y,E)-y'p_y'\big|_{\text{sp}}
\end{equation}
denotes the Legendre-transform of $S$ with respect to $y'$ due to
(\ref{statBed}), and
\begin{equation}
  \hat{\nu} = 
    \begin{cases}
      \nu &   :\left.\frac{\partial^2S}
               {\partial y'{}^2}\right|_{\text{sp}} > 0 \\[.8ex]
      \nu+1 & :\left.\frac{\partial^2S}
               {\partial y'{}^2}\right|_{\text{sp}} < 0
    \end{cases} \;.
\end{equation}

As a general property of Legendre transforms, if we let $u, v$ denote
any of the variables $z,z',y,$ and $E$ which are not involved in the
transformation, we have
\begin{equation}\label{LegTrafo}
  \pabl{\hat{S}}{u} = \pabl{S}{u} \;,\hspace{2cm}
  \pabl{\hat{S}}{p_y'} = -y' \;.
\end{equation}
For the second derivatives, it follows that
\begin{align}
  \ppabl{S}{u}{v} &= \frac{\partial}{\partial u}\pabl{\hat{S}}{v} 
     \notag\\
     &=\ppabl{\hat{S}}{u}{v}+\ppabl{\hat{S}}{p_y'}{v}\pabl{p_y'}{u}
         \Big|_{y'} \displaybreak[0]\\ 
  \intertext{and}
  \ppabl{S}{u}{y'} &= \pabl{p_y'}{u} \notag\\
     &=\frac{\partial}{\partial y'}\pabl{\hat{S}}{u} \notag\\
     &=\ppabl{\hat{S}}{p_y'}{u}\pabl{p_y'}{y'}
      =\ppabl{\hat{S}}{p_y'}{u}\PPabl{S}{y'{}} \;.
\end{align}
Furthermore, we have
\begin{align*}
 D &= \det
      \begin{pmatrix}
        \ppabl{S}{z'}{z} & \ppabl{S}{y'}{z} & \ppabl{S}{E}{z} \\[1ex]
        \ppabl{S}{z'}{y} & \ppabl{S}{y'}{y} & \ppabl{S}{E}{y} \\[1ex]
        \ppabl{S}{z'}{E} & \ppabl{S}{y'}{E} & \PPabl{S}{E}
      \end{pmatrix} \\
   &= \det
      \begin{pmatrix}
        \ppabl{\hat{S}}{z'}{z}+\ppabl{\hat{S}}{p_y'}{z'}\pabl{p_y'}{z}&
        \pabl{p_y'}{z} &
        \ppabl{\hat{S}}{E}{z}+\ppabl{\hat{S}}{p_y'}{E}\pabl{p_y'}{z}
       \\[1ex]
        \ppabl{\hat{S}}{z'}{y}+\ppabl{\hat{S}}{p_y'}{z'}\pabl{p_y'}{y}&
        \pabl{p_y'}{y} &
        \ppabl{\hat{S}}{E}{y}+\ppabl{\hat{S}}{p_y'}{E}\pabl{p_y'}{y}
       \\[1ex]  
        \ppabl{\hat{S}}{z'}{E}+\ppabl{\hat{S}}{p_y'}{z'}\pabl{p_y'}{E}&
        \pabl{p_y'}{E} &
        \PPabl{\hat{S}}{E} +\ppabl{\hat{S}}{p_y'}{E}\pabl{p_y'}{E}
      \end{pmatrix} \;.  
\end{align*}
The second terms in the first and third columns of this matrix are
multiples of the second column, and can thus be omitted. This yields
\begin{equation}\begin{split}
 D &= \det
      \begin{pmatrix}
        \ppabl{\hat{S}}{z'}{z} &
        \PPabl{S}{y'{}}\ppabl{\hat{S}}{p_y'}{z} &
        \ppabl{\hat{S}}{E}{z}
       \\[1ex]
        \ppabl{\hat{S}}{z'}{y} &
        \PPabl{S}{y'{}}\ppabl{\hat{S}}{p_y'}{y} &
        \ppabl{\hat{S}}{E}{y}
       \\[1ex]
        \ppabl{\hat{S}}{z'}{E} &
        \PPabl{S}{y'{}}\ppabl{\hat{S}}{p_y'}{E} &
        \PPabl{\hat{S}}{E}
      \end{pmatrix} \\
   &= \PPabl{S}{y'{}}\,\det
      \begin{pmatrix}
        \ppabl{\hat{S}}{z'}{z} &
        \ppabl{\hat{S}}{p_y'}{z} &
        \ppabl{\hat{S}}{E}{z}
       \\[1ex]
        \ppabl{\hat{S}}{z'}{y} &
        \ppabl{\hat{S}}{p_y'}{y} &
        \ppabl{\hat{S}}{E}{y}
       \\[1ex]
        \ppabl{\hat{S}}{z'}{E} &
        \ppabl{\hat{S}}{p_y'}{E} &
        \PPabl{\hat{S}}{E}
      \end{pmatrix} \;.
\end{split}\end{equation}
If we denote the remaining determinant by $\hat{D}$, we obtain from
(\ref{TrGa})
\begin{equation*}
  \Tr G = \frac{1}{2\pi im\hbar^2} \int dy\,dz\,dp_y'
    \sqrt{|\hat{D}|}\,
    \exp\left\{\frac{i}{\hbar}\left(\hat{S}+yp_y'\right)
               -i\frac{\pi}{2}\hat{\nu}\right\}
    \Bigg|_{z'=z+2\pi m} \;.
\end{equation*}

From our choice of $z$ along the periodic orbit, we have
\begin{equation*}
  \pabl H{\vec p} = \dot{\vec x} = (\dot z, 0) \;.
\end{equation*}
Taking derivatives of the Hamilton-Jacobi equations
\begin{equation*}
  H\left(\pabl S{\vec x'}, \vec x'\right) = E \;, \qquad\qquad
  H\left(-\pabl S{\vec x}, \vec x\right) = E
\end{equation*}
with respect to $z, z'$ and $y$ and using (\ref{LegTrafo}), we therefore
obtain
\begin{gather*}
  \ppabl {\hat S}{z'}z = \ppabl{\hat S}{p_y'}z 
                       = \ppabl{\hat S}{z'}y  = 0 \;, \\
  \ppabl{\hat S}{z'}E = \frac{1}{\dot z'} \;, \qquad\qquad
  \ppabl{\hat S}Ez = \frac{1}{\dot z} \;,
\end{gather*}
so that
\begin{equation}
  \hat{D} = \frac{1}{\dot{z}\dot{z}'}\ppabl{\hat{S}}{y}{p_y'} \;.
\end{equation}
Using this relation, the $z$-integration can trivially be
performed. It yields the time consumed during one cycle. This time
depends on the coordinates $y$ and $p_y$ and is different in general
from the orbital period of the central periodic orbit. We denote it by
$\pabl{\hat{S}}{E}(y,p_y')$:
\begin{equation}
  \Tr G = \frac{1}{2\pi im\hbar^2} \int dy\,dp_y' \pabl{\hat{S}}{E}
    \sqrt{\left|\ppabl{\hat{S}}{y}{p_y'}\right|}\,
    \exp\left\{\frac{i}{\hbar}\left(\hat{S}+yp_y'\right)
        -i\frac{\pi}{2}\hat{\nu}\right\} \;.
\end{equation}
Finally, we obtain the contribution of the orbits under study to the
density of states:
\begin{equation}\label{IntDarst}\begin{split} 
 d(E) &= -\frac{1}{\pi}\,\Im\Tr G \\
   &= \frac{1}{2\pi^2m\hbar^2} \,\Re \int dy\,dp_y' \pabl{\hat{S}}{E}
    \sqrt{\left|\ppabl{\hat{S}}{y}{p_y'}\right|} \\
   &\hspace{4cm}
      \times
      \exp\left\{\frac{i}{\hbar}\left(\hat{S}+yp_y'\right)
        -i\frac{\pi}{2}\hat{\nu}\right\} \;.
\end{split}\end{equation}

The exponent function
\begin{equation}\label{fDef}
  f(y,p_y',E) := \hat{S}(y,p_y',E) +yp_y'
\end{equation}
has to be related to known functions. The only information on $f$
we possess is the distribution of its stationary points: They
correspond to classical periodic orbits.

The classification of real-valued functions with respect to the
distribution of their stationary points is achieved within the
mathematical framework of \emph{catastrophe theory} \cite{Pos78}. The
object of study there are families of functions $\Phi(\vec{x}, Z)$
depending on $k$ so-called \emph{state variables} $\vec{x}$ and
indexed by $r$ \emph{control variables} $Z$. For any fixed control
$Z$, the function $\Phi(\vec{x},Z)$ is assumed to have a stationary
point at the origin and to take 0 as its stationary value
there. Further stationary points may or may not exist in a
neighborhood of the origin. As the control parameters are varied, such
additional stationary points may collide with the central stationary
point, they may be born or destroyed. The aim of catastrophe theory is
to qualitatively understand how these bifurcations of stationary
points can take place. More precisely, two families $\Phi_1$ and
$\Phi_2$ of functions as described above are regarded as eqivalent if
there is a diffeomorphism $\psi_c$ of control space and a
control-dependent family $\psi_s(Z)$ of diffeomorphisms of state space
which keep the origin fixed, such that
\begin{equation}
  \Phi_2(\vec{x},Z) = \Phi_1(\psi_s(\vec{x},Z), \psi_c(Z)) \;.
\end{equation}
Equivalence classes with respect to this relation are known as
catastrophes.
Catastrophes having a codimension of at most four, that is,
catastrophes which can generically be observed if no more than four
control parameters are varied, have completely been classified by
Thom. They are known as the seven \emph{elementary catastrophes}. Each
of these catastrophes can be represented by a polynomial in one or two
variables.

In our discussion of periodic orbits the energy serves as the only control
parameter. However, we are only interested in stationary points of
functions which can be obtained as action functions in Hamiltonian systems.
Due to this restriction, we can generically observe scenarios which would
have higher codimensions in the general context of catastrophe theory, so
that catastrophes of codimension greater than one are relevant for our
purpose. The variation of energy then defines a path $Z(E)$ in an abstract
higher-dimensional control space.

In earlier work on the contruction of uniform approximations close to
nongeneric bifurcations, Main and Wunner \cite{Mai97a,Mai98a}
succeeded in relating the action function describing the bifurcation
scenario to one of the elementary catastrophes. In our case, however,
this approach fails because the codimension (in the sense of
catastrophe theory) of the action function is even higher than
four. Nevertheless, we can make use of the equivalence relation of
catastrophe theory, because, as was shown above, the normal form has
got stationary points which exactly correspond to the periodic orbits
of the classical system. This observation enables us in principle to
systematically construct ansatz functions for any bifurcation scenario
encountered in a Hamiltonian system using normal form theory. We are
thus led to making the ansatz
\begin{equation}\label{KatastAnsatz}
  f(y,p_y';E) = S_0(E) + \Phi(\psi_s(y,p_y';E),\psi_c(E)) \;.
\end{equation}
Here, the energy $E$ serves as the control parameter, $\Phi$ denotes
the normal form of the bifurcation scenario, $\psi_s$ and $\psi_c$
unknown coordinate changes as in the general context of catastrophe
theory, and $S_0(E)$ is the action of the central periodic orbit,
which has to be introduced here to make both sides equal at the
origin. The unknown transformations $\psi_s$ and $\psi_c$ can easily
be accounted for because they can only manifest themselves in
appropriate choices of the free parameters occuring in the normal
form. Inserting the ansatz (\ref{KatastAnsatz}) into (\ref{IntDarst})
and transforming the integration measure to new coordinates $(Y,P_Y')
= \psi_s(y,p_y';E)$, we obtain
\begin{multline*}
  d(E) = \frac{1}{2\pi^2m\hbar^2} \,\Re
    \exp\left\{\frac{i}{\hbar}S_0(E)-i\frac{\pi}{2}\hat{\nu}\right\}\\ 
   \times
   \int dYdP_Y' \pabl{\hat{S}}E
     \sqrt{\left|\ppabl{\hat{S}}{y}{p_y'}\right|}
                                      \frac{1}{|\det\Jac\psi_s|}
     \exp\left\{\frac{i}{\hbar}\Phi(Y,P_Y')\right\} \;,
\end{multline*}
where $\Jac\psi_s$ denotes the Jacobian matrix of $\psi_s$ with
respect to the variables $y$ and $p_y'$.
Differentiating the ansatz (\ref{KatastAnsatz}) twice, we get the
matrix equation
\begin{equation*}
  \frac{\partial^2f}{\partial(y,p_y')^2} = (\Jac\psi_s)^T
  \frac{\partial^2\Phi}{\partial(Y,P_Y')^2} \Jac\psi_s \;,
\end{equation*}
so that the determinants satisfy
\begin{equation}
  |\det\Jac\psi_s| = \sqrt{\frac{|\Hess f|}{|\Hess \Phi|}} \;,
\end{equation}
and the density of states finally reads
\begin{multline}\label{unifInt}
  d(E) = \frac{1}{2\pi^2m\hbar^2} \,\Re
    \exp\left\{\frac{i}{\hbar}S_0(E)-i\frac{\pi}{2}\hat{\nu}\right\}\\ 
   \times
   \int dYdP_Y' \pabl{\hat{S}}E
     \sqrt{\left|\ppabl{\hat{S}}{y}{p_y'}\right|}
     \sqrt{\frac{|\Hess \Phi|}{|\Hess f|}}
     \exp\left\{\frac{i}{\hbar}\Phi(Y,P_Y')\right\} \;.
\end{multline}

The exponent function in the integrand of the remaining integral is
given by the normal form describing the bifurcation scenario, which
 was calculated in the preceding section for the present case. The
normal form parameters, however, still have to be determined. On the
other hand, the coefficient
\begin{equation}\label{XDef}
  X := \pabl{\hat{S}}E
     \sqrt{\left|\ppabl{\hat{S}}{y}{p_y'}\right|}
     \sqrt{\frac{|\Hess \Phi|}{|\Hess f|}}
\end{equation}
is completely unknown. To evaluate (\ref{unifInt}), we have to
establish a connection between $X$ and classical periodic orbits. As
periodic orbits correspond to stationary points of the normal form, we
will now analyse the behavior of $X$ at stationary points of the
exponent.

By (\ref{fDef}), the Hessian matrix of $f$ is given by
\begin{equation*}
  \frac{\partial^2 f}{\partial (y,p_y')^2} =
    \begin{pmatrix}
      \PPabl{\hat{S}}{y} & \ppabl{\hat{S}}{y}{p_y'}+1 \\[1ex]
      \ppabl{\hat{S}}{y}{p_y'}+1 & \PPabl{\hat{S}}{p_y'{}}
    \end{pmatrix}
\end{equation*}
so that the Hessian determinant reads
\begin{equation*}\begin{split}
  \Hess f&= \PPabl{\hat{S}}{y}\PPabl{\hat{S}}{p_y'{}}
            -\left(\ppabl{\hat{S}}{y}{p_y'} +1\right)^2\\
         &= -\left(1+\left(\ppabl{\hat{S}}{y}{p_y'}\right)^2-
               \PPabl{\hat{S}}{y}\PPabl{\hat{S}}{p_y'{}}\right)
               -2\ppabl{\hat{S}}{y}{p_y'} \;.
\end{split}\end{equation*}
As can be shown, in a two-degree-of-freedom system the monodromy matrix of
a periodic orbit can be expressed in terms of the action function as
\begin{equation}\label{TrMS}
  \Tr M =
    -\left(\ppabl{\hat{S}}{y}{p_y'{}}\right)^{-1}
    \left\{1+\left(\ppabl{\hat{S}}{y}{p_y'{}}\right)^2
            -\PPabl{\hat{S}}{y}\PPabl{\hat{S}}{p_y'{}}\right\} \;,
\end{equation}
so that
\begin{equation}
  \Hess f \stackrel{\text{sp}}{=} (\Tr M -2)\ppabl {\hat{S}}{y}{p_y'}
\end{equation}
and
\begin{equation}
  \sqrt{\frac{1}{|\Hess f|}}\sqrt{\left|\ppabl{\hat{S}}{y}{p_y'}\right|}
  \stackrel{\text{sp}}{=} \frac{1}{\sqrt{|\Tr M-2|}} \;.
\end{equation}
Furthermore, we make use of the fact that at a stationary point the
derivative $\partial\hat S/\partial E$ gives the orbital period of the
corresponding periodic orbit. For the central periodic orbit, this is
$m$ times the primitive period $mT_0$, for a satellite orbit, however,
it gives a single primitive period $T_s$. Altogether, these results
yield
\begin{equation}\label{XStat}
  X\stackrel{\text{sp}}{=}\frac{\{m\}T}{\sqrt{|\Tr M-2|}}
    \sqrt{|\Hess \Phi|}\;,
\end{equation}
where the notation $\{m\}$ is meant to indicate that the factor of $m$
has to be omitted for a satellite orbit.  This expression can be
calculated once the normal form parameters have suitably been
determined.

Furthermore, (\ref{XStat}) allows us to check that (\ref{unifInt})
does indeed reduce to Gutz\-wil\-ler's isolated-orbits contributions
if the distance from the bifurcations is large: If the stationary
points of the normal form are sufficiently isolated, we can return to
a stationary-phase approximation of the integral.  We will first
calculate the contribution of the stationary point at $Y=P_Y'=0$,
which corresponds to the central periodic orbit. If we use $\Phi(0)=0$
and (\ref{XStat}) and let $\lambda$ denote the number of negative
eigenvalues of the Hessian matrix
$\left.\frac{\partial^2\Phi}{\partial(Y',P_Y)^2}\right|_0$, this
contribution reads
\begin{equation*}\begin{split}
  &\quad  \frac{1}{2\pi^2m\hbar^2} \frac{mT_0}{\sqrt{\Tr M_0-2}}
    \sqrt{|\Hess\Phi|\big|_0} \,
    \Re\exp\left\{\frac{i}{\hbar}S_0-i\frac{\pi}{2}\hat{\nu}\right\}\,
    (2\pi i\hbar)
    \frac{\exp\left\{-i\frac{\pi}{2}\lambda\right\}}
       {\sqrt{|\Hess\Phi|\big|_0}} \\
 &= \frac{1}{\pi\hbar}\frac{T_0}{\sqrt{\Tr M_0-2}}\,
    \Re\exp\left\{\frac{i}{\hbar}S_0
      -i\frac{\pi}{2}\left(\hat{\nu}+\lambda-1\right)\right\} \;.
\end{split}\end{equation*}
If we identify $\hat{\nu}+\lambda-1$ with the Maslov index of the
central periodic orbit, and note that in a two-degree-of-freedom system
$\Tr M -2 = \det(M-I)$, this is just Gutzwiller's periodic-orbit
contribution.

Satellite orbits contribute only at energies where they are real. In
this case, every satellite orbit corresponds to $m$ stationary points
of the normal form. Altogether, they contribute
\begin{equation*}\begin{split}
 &\quad m\frac{1}{2\pi^2m\hbar^2} \frac{T_s}{4\sqrt{\Tr M_s-2}}
    \sqrt{|\Hess\Phi|\big|_s} \\
 &\qquad\qquad 
    \Re\exp\left\{\frac{i}{\hbar}S_0-i\frac{\pi}{2}\hat{\nu}\right\}\,
    (2\pi i\hbar)\exp\left\{\frac{i}{\hbar}\Phi_s\right\}
    \frac{\exp\left\{-i\frac{\pi}{2}\lambda'\right\}}
       {\sqrt{|\Hess\Phi|\big|_s}} \\
 &= \frac{1}{\pi\hbar}\frac{T_s}{\sqrt{\Tr M_s-2}}
    \Re\exp\left\{\frac{i}{\hbar}S_s
      -i\frac{\pi}{2}\left(\hat{\nu}+\lambda'-1\right)\right\}\;.
\end{split}\end{equation*}
Here, the number of negative eigenvalues of
$\left.\frac{\partial^2\Phi}{\partial(Y',P_Y)^2}\right|_s$ was denoted by
$\lambda'$, and we made use of the fact that, according to our ansatz
(\ref{KatastAnsatz}), $S_0(E)+\Phi_s$ equals the action $S_s(E)$ of the
satellite orbit. Thus, we also obtain Gutzwiller's contribution for
satellite orbits, provided that $\hat{\nu}+\lambda'-1$ is the Maslov index
of the satellite orbit. We can regard this as a consistency condition
which allows us to calculate the difference in Maslov index between the
central periodic orbit and the satellites from the normal form.

Now that we have convinced ourselves that the integral formula
(\ref{unifInt}) is correct, we can go over to its numerical
evaluation. This can be done to different degrees of approximation,
and we are going to present two different approximations in the
following sections.

\subsection{Local Approximation}
\label{locAppSec}

To obtain the simplest approximation possible, we can try to determine
the normal form parameters $a, b, c$ so that the stationary values
(\ref{NF3Wirk}) of the normal form (\ref{NF3Var}) globally reproduce
the actions of the periodic orbits as well as possible. In the spirit
of the stationary-phase method we can further assume the integral in
(\ref{unifInt}) to be dominated by those parts of the coordinate plane
where the stationary points of the exponent are located. As the
uniform approximation is only needed close to a bifurcation, where the
stationary-phase approximation fails, all stationary points lie in a
neighborhood of the origin $I=0$. Thus, we can approximate the
derivative $\partial\hat S/\partial E$ by its value at the origin,
that is, the orbital period $mT_0$ of the central periodic
orbit. Furthermore, we can try and approximate the quotient $(\Tr
M-2)/\Hess\Phi$ by a constant $k$. We then obtain
\begin{multline}\label{lokInt}
  d(E)\approx\frac{1}{2\pi^2\hbar^2}\frac{T_0}{\sqrt{|k|}} \,
    \Re \exp\left\{\frac{i}{\hbar}S_0(E)-i\frac{\pi}{2}\hat{\nu}
       \right\} \\
    \times
    \int dYdP_Y' \exp\left\{\frac{i}{\hbar}\Phi(Y,P_Y')\right\} \;.
\end{multline}
Thus, the density of states is approximately given by the integral of
a known function which can be evaluated numerically.

\begin{figure}
  \wlabelgfx{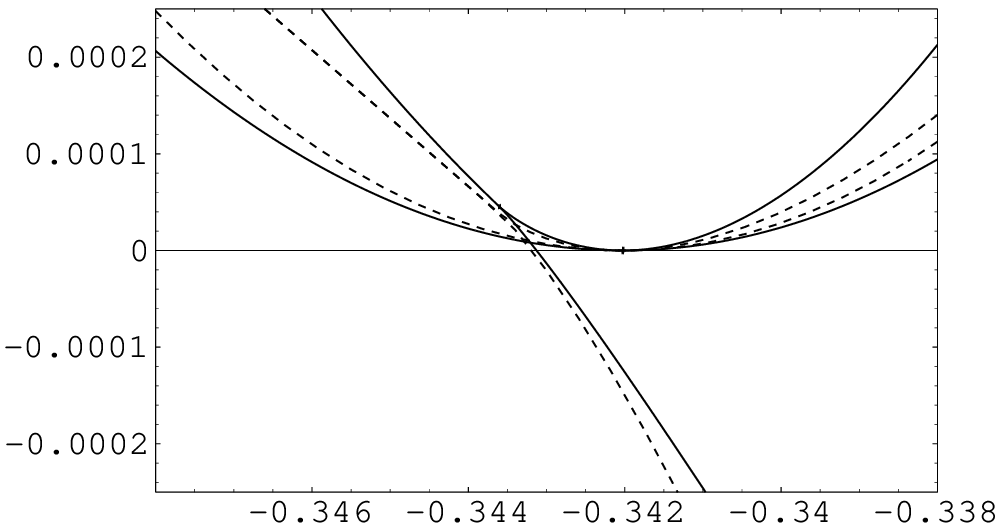}
     {\mbox{\large $\mathsf{\tilde{E}}$}}
     {\mbox{\large $\mathsf{Re\,\Delta\tilde{S}/2\pi}$}}{12cm}\\[3ex]
  \wlabelgfx{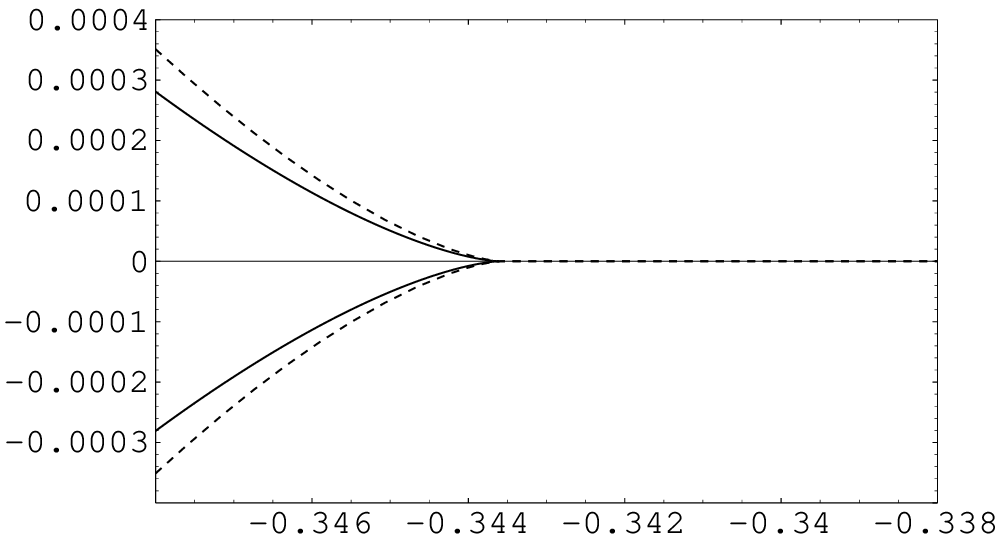}
     {\mbox{\large $\mathsf{\tilde{E}}$}}
     {\mbox{\large $\mathsf{Im\,\Delta\tilde{S}/2\pi}$}}{12cm}                    \caption{\label{JVglAbb}
    Comparison of the action differences calculated from the normal form
    (\ref{NF3Var}) with parameter values (\ref{NF3Params}) to the actual
    action differences. Solid curves: numerically calculated values, dashed
    curves: stationary values of the normal form.}
\end{figure}

\begin{figure}
  \wlabelgfx{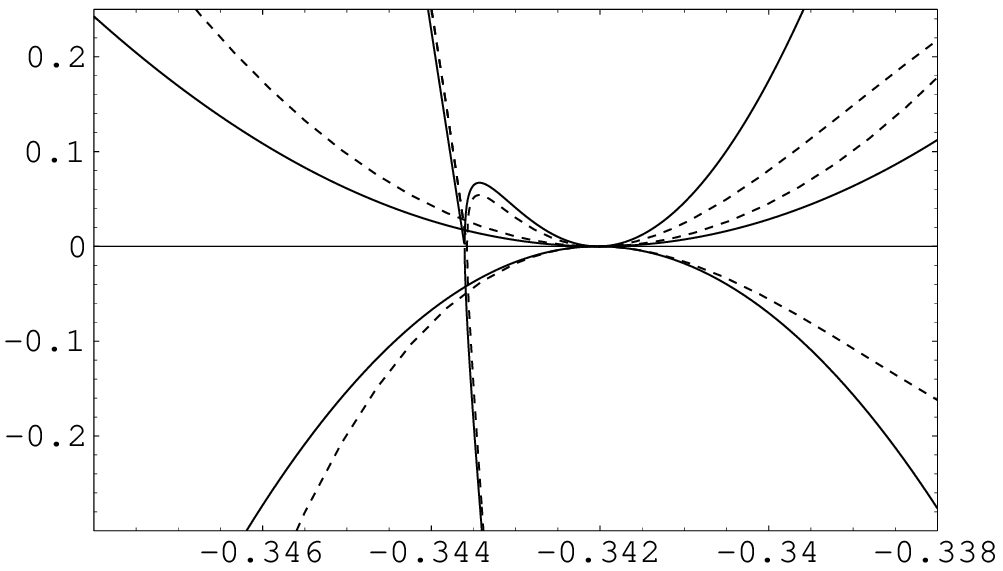}
     {\mbox{\large $\mathsf{\tilde{E}}$}}
     {\mbox{\large $\mathsf{Re\,(Tr\;M-2)}$}}{12cm}\\[3ex]
  \wlabelgfx{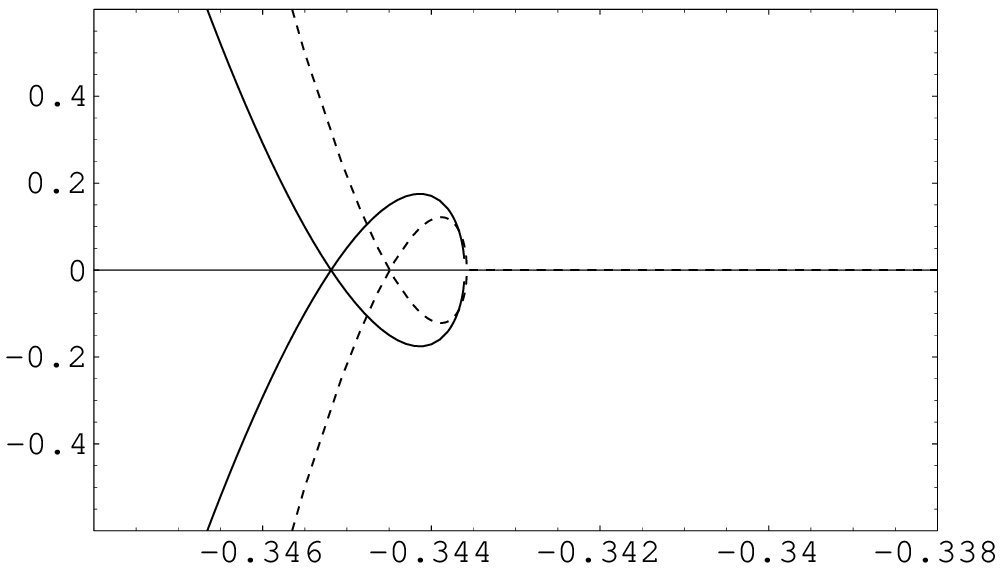}
     {\mbox{\large $\mathsf{\tilde{E}}$}}
     {\mbox{\large $\mathsf{Im\,(Tr\;M-2)}$}}{12cm}                               \caption{\label{DetVglAbb}
    Comparison of the traces of the monodromy matrices calculated from the
    normal form (\ref{NF3Var}) with parameter values (\ref{NF3Params}) to
    the actual traces. Solid curves: numerically calculated traces, dashed
    curves: Hessian determinants of the normal form.}
\end{figure}

In our case, the distance from the period-quadrupling bifurcation is
described by the normal form parameter $\varepsilon$. We choose to
measure this distance by the difference in scaled energy
\begin{equation}\label{epsDef}
  \varepsilon = \tilde{E}-\tilde{E}_c \;.
\end{equation}
Then, to globally reproduce the numerically calculated actions, we use
the parameter values
\begin{equation}\label{NF3Params}\begin{aligned}
  \tilde{a}&=-0.029 \;,\\
  \tilde{b}&=+0.007 \;,\\ 
  \tilde{c}&=-0.052 \;,\\
  \tilde{k}&=11000 \;,
\end{aligned}\end{equation}
where the tilde indicates that the parameters have been adjusted to
the scaled actions $\tilde S/2\pi$. As can be seen from figures
\ref{JVglAbb} and \ref{DetVglAbb}, the action differences and Hessian
determinants calculated from the normal form do indeed qualitatively
reproduce the acual data, although quantitatively the agreement is not
very good. Nevertheless, we will try and calculate the density of
states within the present approximation.

If we are actually going to calculate spectra for different values of
the magnetic field strength, we have to determine the action $S=2\pi w
\cdot (\tilde{S}/2\pi)$ according to the scaling
prescription (\ref{WirkSkal}) with the scaling parameter
$w=\gamma^{-1/3}$. As can easily be seen with the help of
(\ref{NF3Wirk}), this scaling can be achieved by scaling the normal
form parameters according to
\begin{equation}\label{ParamSkal}\begin{aligned}
  a &= \tilde{a}/2\pi w \;,\\
  b &= \tilde{b}/2\pi w \;,\\
  c &= \tilde{c}/4\pi^2w^2 \;.
\end{aligned}\end{equation}
According to its definition (\ref{epsDef}), the parameter
$\varepsilon$ does not scale. Neither does $k$ scale with the magnetic
field strength, because it is given by a quotient of two non-scaling
quantities, but the factor of $2\pi$ in $\Hess\Phi$ has to be taken
into account:
\begin{equation}
  k = \tilde{k}/2\pi \;.
\end{equation}

\begin{figure}
  \vspace{18.0cm}
  \includegraphics{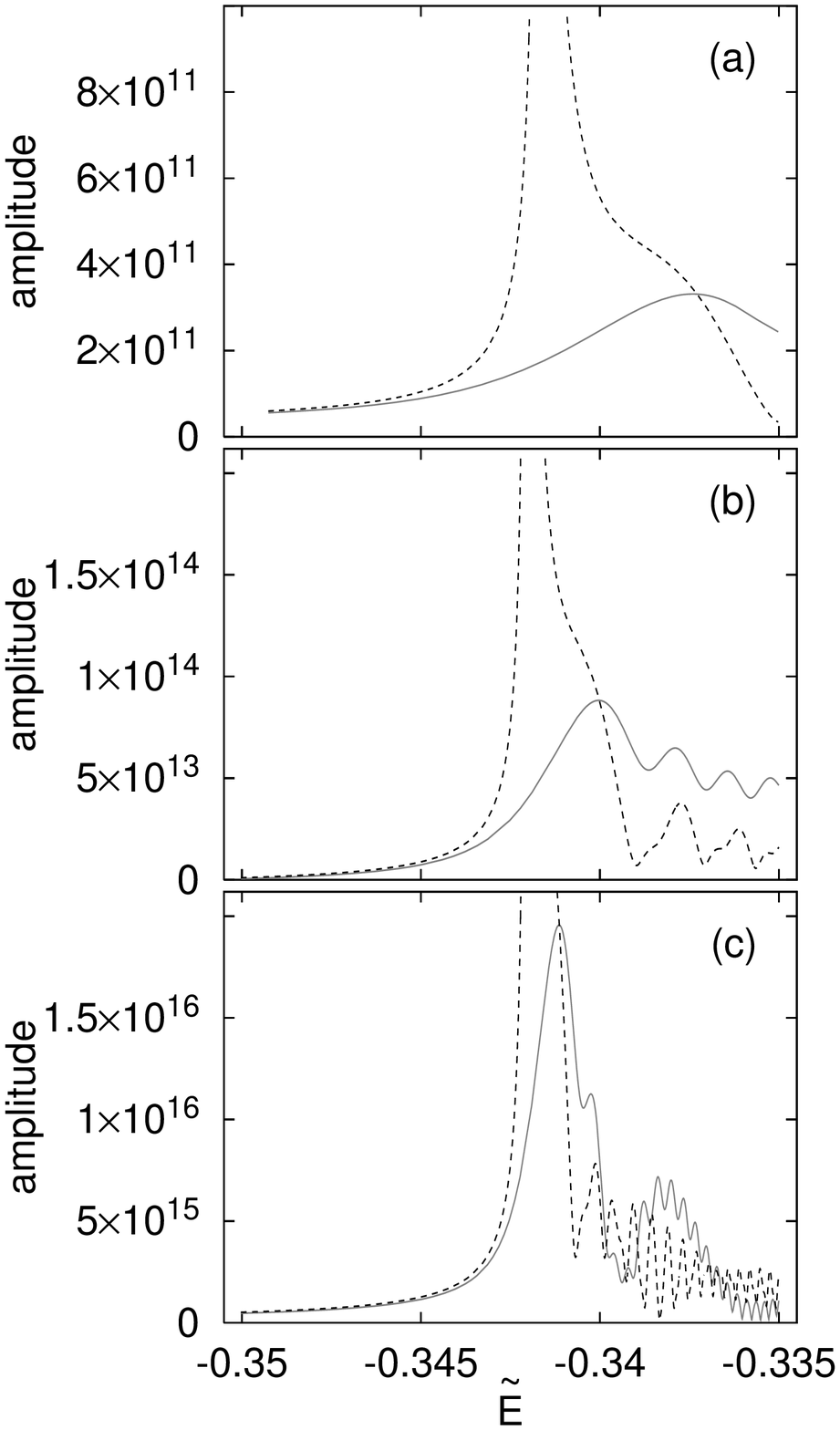}
  \caption{\label{lokAbb}
    Local approximation to the contribution of the considered
    bifurcation scenario to the density of states for three different
    values of the magnetic field strength: (a) $\gamma=10^{-10}$, (b)
    $\gamma = 10^{-12}$, (c) $\gamma=10^{-14}$. Solid curves: local
    approximation, dashed curves: Gutzwiller's trace formula.}
\end{figure}

The local approximation calculated with these data is shown in figure
\ref{lokAbb} for three different values of the magnetic field strength.
Instead of the real part, we actually plotted the absolute value of the
expression in (\ref{lokInt}) to suppress the highly oscillatory factor
$\exp\left\{\frac{i}{\hbar}S_0(E)\right\}$. As was to be expected, the
approximation does indeed give finite values at the bifurcation points, but
does not reproduce the results of Gutzwiller's trace formula as the
distance from the bifurcations is increased. This is due to the fact that
the normal form with the parameter values chosen does not reproduce the
actual orbital data very well. In particular, a better description of the
actions is needed to improve the approximation, because asymptotically the
actions occur as phases in Gutzwiller's trace formula, so that, if the
error in phases $w\tilde S$ is not small compared to $2\pi$, the
interference effects between the contributions of different orbits cannot
correctly be described.

Furthermore, we cannot even expect our local approximation to yield
very accurate values at the bifurcation points themselves, because the
normal form parameters were chosen to globally reproduce the orbital
data, so that a local parameter fit designed to describe the immediate
neighborhood of the bifurcations would lead to different results.

\subsection{Uniform Approximation}
To improve our approximation, we can make use of the fact that the
coordinate transformation $\psi_s$ in (\ref{KatastAnsatz}) is
energy-dependent in general, so that the normal form parameters
$a,b,c$ will also depend on energy. We thus have to choose the
parameters so as to reproduce the numerically calculated action
differences for any fixed $\varepsilon=\tilde E-\tilde E_c$. To
achieve this, we have to solve equations (\ref{NF3Wirk})
\begin{equation}\label{NF3Wirk2}\begin{gathered}
  \Phi_\pm = 4(\eta+\delta^2)\left(\delta\pm\sqrt{\eta+\delta^2}\right)
                +2\eta\delta \;, \\
  \Phi_{-1} = -\frac{\varepsilon^2}{4(a-b)} \;,
\end{gathered}\end{equation}
where
\begin{equation}\begin{gathered}
  \eta = -\frac{\varepsilon}{6c^{1/3}} \;, \\
  \delta = \frac{a+b}{6c^{2/3}} \;,
\end{gathered}\end{equation}
for $a,b,c$. To this end, we introduce
\newcommand{\h}{h}
\begin{equation}\label{hpmDef}\begin{split}
  \h_+ &=\frac{\Phi_++\Phi_-}{8} \\
       &=\delta\left(\eta+\delta^2\right)+\frac{1}{2}\eta\delta \;, \\
  \h_- &=\frac{\Phi_+-\Phi_-}{8} \\
       &=(\eta+\delta^2)^{3/2} \;.
\end{split}\end{equation}
The second equation yields
\begin{equation}\label{etaVonDelta}
  \eta = \h_-^{2/3}-\delta^2 \;.
\end{equation}
Inserting this into the first equation of (\ref{hpmDef}), we obtain
\begin{equation}\label{deltaGlg}
  \delta^3-3\h_-^{2/3}\delta+2\h_+ = 0 \;.
\end{equation}
This is a cubic equation for $\delta$. Its solutions read, from
Cardani's formula,
\begin{equation}\label{deltaLsg}
  \delta  = \frac{\lambda}{2}
              \cbrt
                   {-\left(\sqrt{\Phi_+}+\sqrt{\Phi_-}\right)^2}
           +\frac{\lambda^\ast}{2}
              \cbrt
                {-\left(\sqrt{\Phi_+}-\sqrt{\Phi_-}\right)^2}\;,
\end{equation}
where $\lambda\in\{1,-\frac{1}{2}\pm i\frac{\sqrt{3}}{2}\}$ is a cube
root of unity. If the discriminant
\begin{equation}\label{Diskr}
  D = \h_+^2 -\h_-^2 = \frac{1}{16}\Phi_+\Phi_- \;,
\end{equation}
is positive, there is only one real solution for $\delta$, which has
$\lambda=1$. If, however, $D<0$, all three solutions are real. In this
case we have to choose one solution before we can proceed.

Using the correspondence between stationary points and periodic orbits
discussed above, we find from figure \ref{DataFig} that $\Phi_+>0$,
and there exists an $\varepsilon_0<0$ so that $\Phi_-<0$ if
$\varepsilon>\varepsilon_0$ and $\Phi_->0$ if
$\varepsilon<\varepsilon_0$. Thus, from (\ref{Diskr}), we have
$D\lessgtr 0$ if $\varepsilon\gtrless\varepsilon_0$, and we have to
choose $\lambda=1$ if $\varepsilon<\varepsilon_0$ to make $\delta$
real.

\renewcommand{\ord}[1]{{\cal O}\left(\varepsilon^{#1}\right)}
\newcommand{\oo}{\ord{2}}
To determine the correct choice of $\lambda$ for
$\varepsilon>\varepsilon_0$, we demand that $\delta$ must depend on
$\varepsilon$ continuously. Thus, $\lambda$ can only change at
energies where (\ref{deltaGlg}) has a double root, viz.~ $D=0$ or
$\varepsilon\in\{0,\varepsilon_0\}$. Therefore, it suffices to
determine $\lambda$ in a neighborhood of $\varepsilon=0$.
Close to $\varepsilon=0$, the action differences can be seen from
figure \ref{DataFig} to behave like
\begin{align*}
  \Phi_+ &= \alpha^2\varepsilon^2+\ord{3} \;,\\
  \Phi_- &= -\Gamma-\beta\varepsilon+\oo
\end{align*}
with positive real constants $\alpha, \beta, \Gamma$. Equations
(\ref{deltaLsg}) and (\ref{etaVonDelta}) then allow us to expand
$\eta$ in a Taylor series in $\varepsilon$:
\begin{multline}
 \eta = \left(\frac{1}{4}-(\Re\lambda)^2\right)\Gamma^{2/3} \\
      +\left(\left(\frac{1}{4}-(\Re\lambda)^2\right)
             \frac{2\beta}{3\Gamma^{1/3}}
            -\Re\lambda\, \Im\lambda\,
             \frac{4\alpha\sign\varepsilon
                               \sqrt{\Gamma}}{3\Gamma^{1/3}}
      \right)\varepsilon + \oo
\end{multline}
If we require this result to reproduce the definition
\begin{equation*}
  \eta = -\frac{1}{6c^{1/3}}\varepsilon,\qquad\qquad
  -\frac{1}{6c^{1/3}}>0 \;,
\end{equation*}
we find the conditions
\begin{gather*}
  \Re\lambda = -\frac{1}{2} \;,\\
  \Im\lambda\,\frac{2\alpha\sigma}{3\Gamma^{1/3}} > 0 \;;
\end{gather*}
and are thus led to the correct choices of $\lambda$:
\begin{equation}\label{lambdaWahl}
  \lambda =
    \begin{cases}
      1 & : \Phi_->0 \\
      -\frac{1}{2}+i\frac{\sqrt{3}}{2}\sign\varepsilon & : \Phi_-<0
    \end{cases} \;.
\end{equation}
Using this result, we obtain $\delta$ and $\eta$ from (\ref{deltaLsg})
and (\ref{etaVonDelta}). Finally, from (\ref{EtaDeltaDef}) and
(\ref{NF3Wirk2}) we can explicitly determine the normal form
parameters $a, b, c$ as functions of the energy $\varepsilon$ and the
action differences $\Phi_+,\Phi_-,$ and $\Phi_{-1}$:
\begin{equation}\label{abc}\begin{gathered}
  c=-\left(\frac{\varepsilon}{6\eta}\right)^3 \;,\\
  a+b = 6 c^{2/3} \delta\;, \qquad\qquad
     a-b=-\frac{\varepsilon}{4\Phi_{-1}} \;,\\ 
  a = 3c^{2/3}\delta - \frac{\varepsilon}{8\Phi_{-1}}\;, \qquad\qquad
     b = 3c^{2/3}\delta + \frac{\varepsilon}{8\Phi_{-1}} \;.
\end{gathered}\end{equation}

Now that the normal form $\Phi$ has been completely specified, a
suitable approximation to the coefficient $X$ remains to be found. We
shall assume $X$ to be independent of the angular coordinate
$\varphi$, and because, from (\ref{XStat}), the value of $X$ is known
at the stationary points of $f$ at four different values of $I$
(including $I=0$), we can approximate $X$ by the third-order
polynomial $p(I)$ interpolating between the four given points, so that
our uniform approximation takes its final form
\begin{multline}\label{unifLsg}
  d(E) = \frac{1}{2\pi^2m\hbar^2}\,\Re \exp
     \left\{\frac{i}{\hbar}S_0(E)-i\frac{\pi}{2}\hat{\nu}\right\} \\
     \times
     \int dY\,dP_Y'p(I) \exp\left\{\frac{i}{\hbar}\Phi(Y,P_Y')\right\}
  \;,
\end{multline}
This choice insures that our approximation reproduces Gutzwiller's
isolated-orbits formula if, sufficiently far away from the
bifurcation, the integral is evaluated in stationary-phase
approximation. Thus, our solution is guaranteed to exhibit the correct
asymptotic behavior. On the other hand, as was shown above, very close
to the bifurcations the integral is dominated by the region around the
origin. As our interpolating polynomial assumes the correct value of
$X$ at $I=0$, we can expect our uniform approximation to be very
accurate in the immediate neighborhood of the bifurcations, too, and
thus to yield good results for the semiclassical density of states in
the complete energy range.

\begin{figure}
  \vspace{18.0cm}
  \includegraphics{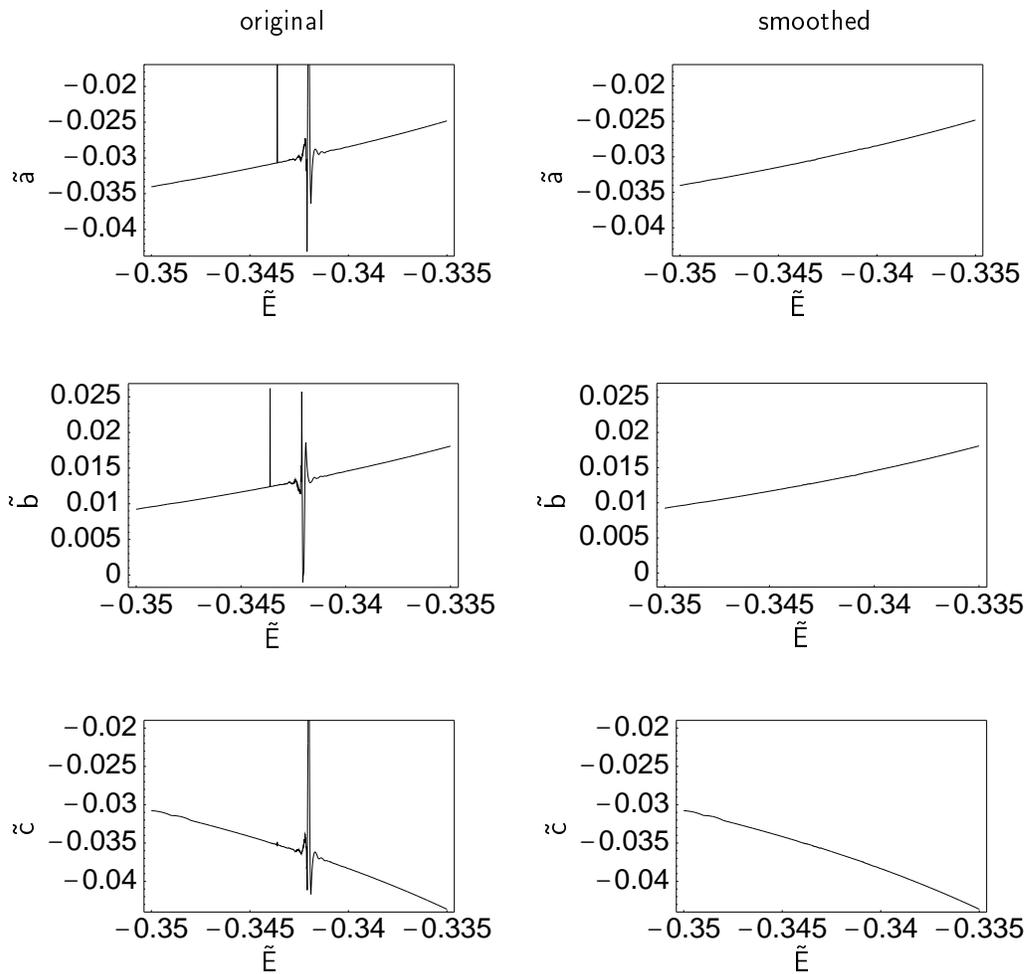}
  \caption{\label{abcAbb}
    Values of the normal form parameters calculated from (\ref{abc}) which
    were used for the uniform approximation.}
\end{figure}

The values of the normal form parameters $a,b,c$ calculated from
(\ref{abc}) are shown in figure \ref{abcAbb}. Obviously, their
calculation becomes numerically unstable close to the
bifurcations. This is due to two reasons:
\begin{itemize}
\item
  As input data to (\ref{abc}), we need action differences between the
  central orbit and the satellite orbits. Close to the bifurcations,
  these differences become arbitrarily small and can thus be
  determined from the numerically calculated actions to low precision
  only.

\item
  The parameter $c$ is given by the quotient of $\eta$ and
  $\varepsilon$, which quantities both vanish at the bifurcation
  energy. As the bifurcation energy $\tilde E_c$, and hence
  $\varepsilon$, is not known to arbitrarily high precision, the
  zeroes of the numerator and the denominator do not exactly coincide,
  so that the quotient assumes a pole.
\end{itemize}
We can smooth the parameters by simply interpolating their values from
the numerically stable to the unstable regimes. As the dependence of
the parameters on energy is very smooth, we can expect this procedure
to yield accurate results.

\begin{figure}
  \vspace{18.0cm}
  \includegraphics{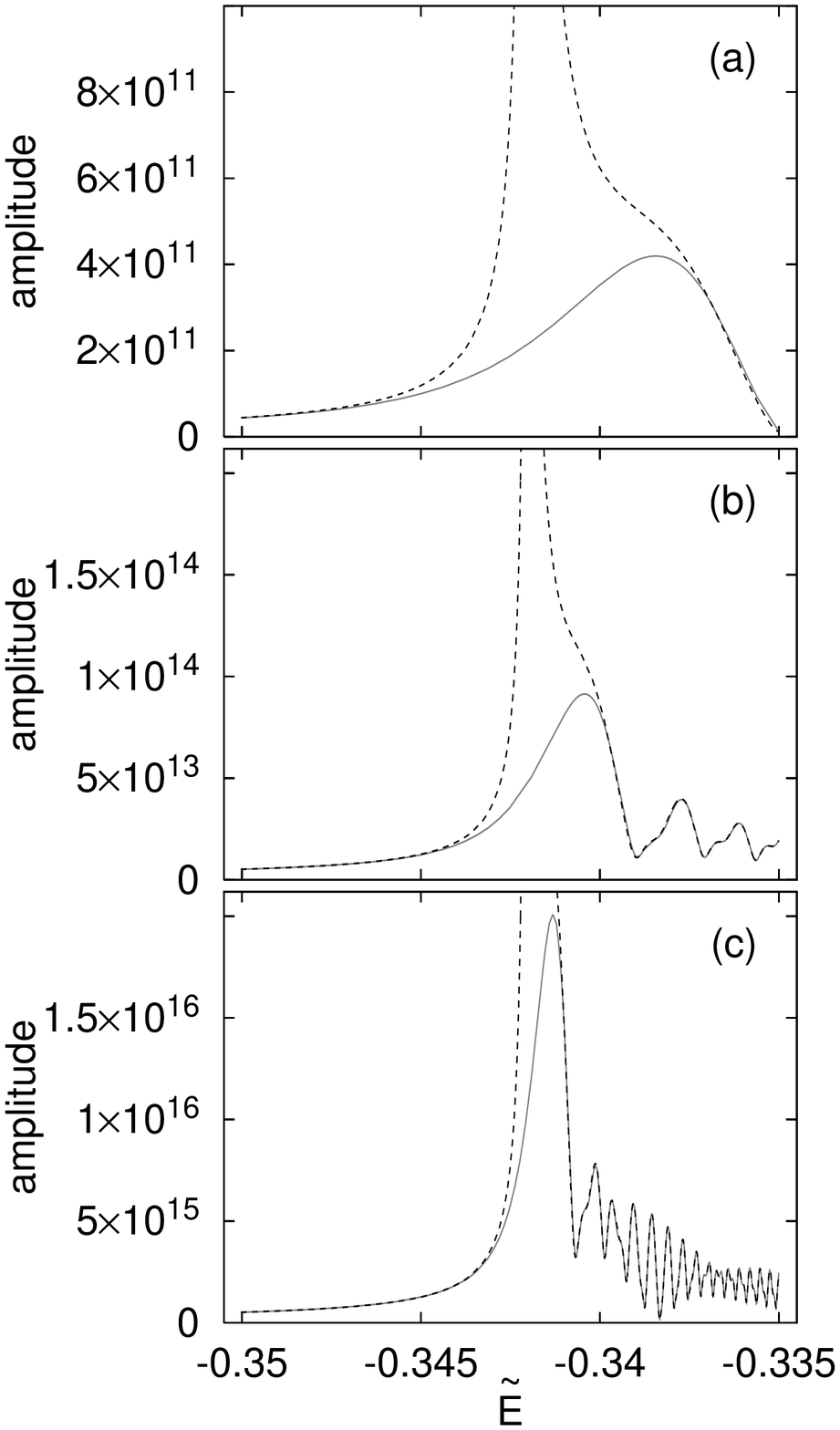}
  \caption{\label{unifAbb}
    Uniform approximation to the contribution of the considered
    bifurcation scenario to the density of states for three different
    values of the magnetic field strength: (a) $\gamma=10^{-10}$, (b)
    $\gamma = 10^{-12}$, (c) $\gamma=10^{-14}$. Solid curves: Uniform
    approximation, dashed curves: Gutzwiller's trace formula.}
\end{figure}

The uniform approximation was calculated for the same values of the
magnetic field strength as was the local approximation. The results
are displayed in figure \ref{unifAbb}. They are finite at the
bifurcation energies and do indeed reproduce the results of
Gutzwiller's trace formula as the distance from the bifurcation
increases. Even the complicated oscillatory structures in the density
of states which are caused by interferences betweeen contributions
from different periodic orbits are perfectly reproduced by our uniform
approximation. We can also see that the higher the magnetic field
strength, the farther away from the bifurcation energies is the
asumptotic (Gutzwiller) behavior acquired. In fact, for the largest
field strength $\gamma=10^{-10}$ the asymptotic regime is not reached
at all in the energy range shown. This behavior can be traced back to
the fact that, due to the scaling properties of the Diamagnetic Kepler
Problem, the scaling parameter $\gamma^{-1/3}$ plays the r\^ole of an
effective Planck's constant, therefore the lower $\gamma$ becomes, the
more accurate the semiclassical approximation will be.

\section{Summary}

We have shown that in generic Hamiltonian systems bifurcations of
ghost orbits can occur besides the bifurcations of real orbits. If
they occur in the neighborhood of a bifurcation of a real orbit, they
produce signatures in semiclassical spectra in much the same way as
bifurcations of real orbits and therefore are of equal importance to a
semiclassical understanding of the quantum spectra. Furthermore, we
have shown that the technique of normal form expansions traditionally
used to construct uniform approximations taking account of
bifurcations of real orbits only can be extended to also include the
effects of ghost orbit bifurcations. Thus, normal form theory offers
techniques which will allow us, at least in principle, to calculate
uniform approximations for arbitrarily complicated bifurcation
scenarios.

The effects ghost orbit bifurcations exert on semiclassical spectra
were illustrated by way of example of the period-quadrupling
bifurcation of the balloon-orbit in the Diamagnetic Kepler
Problem. This example was chosen mainly because of its simplicity,
because the balloon-orbit is one of the shortest periodic orbits in
the Diamagnetic Kepler Problem, and the period-quadrupling is the
lowest period-$m$-tupling bifurcation that can exhibit the
island-chain-bifurcation typical of all higher $m$. We can therefore
expect ghost orbit bifurcations also to occur for longer orbits and in
connection with higher period-$m$-tupling bifurcations. This conjeture
is confirmed by the discussion of the various bifurcation scenarios
described by the higher-order normal form (\ref{NF3}), which reveals
ghost orbit bifurcations close to a period-quadrupling in three out of
four possible cases. Thus, ghost orbit bifurcations will be a common
occurence in generic Hamiltonian systems. Their systematic study
remains an open problem for future work.

\clearpage

\clearpage  


\end{document}